\documentclass[a4paper,twocolumn,amsmath,amssymb,prx,preprintnumbers,notitlepage,preprintfigures,longbibliography]{revtex4-2}

\usepackage{graphicx}\usepackage{dcolumn}\usepackage[tight]{subfigure}
\usepackage{amsmath}
\usepackage{verbatim}
\usepackage[usenames,dvipsnames]{color}
\usepackage{bm} \usepackage{bbm}
\usepackage{natbib}
\usepackage{xspace}
\usepackage{marginnote}
\usepackage{mathtools}
\usepackage{dsfont}
\usepackage{soul} \usepackage{hyperref} \hypersetup{colorlinks=true,linktoc=all,linkcolor=blue,breaklinks=true,citecolor=blue,urlcolor=blue}
\usepackage{csquotes} 
\usepackage{etoolbox} \robustify{\uparrow}
\robustify{\downarrow}
\robustify{\sum}
\robustify{\int}
\robustify{\nonumber}
\robustify{\cite}
\robustify{\footnote}

\renewrobustcmd{\Re}{{\text{Re}}}
\renewrobustcmd{\Im}{{\text{Im}}}

\newrobustcmd{\sdag}{{\bm{\dag}}}  \newrobustcmd{\tot}{\text{tot}} \renewrobustcmd{\l}{\text{L}}   \newrobustcmd{\n}{\text{N}}   
\newrobustcmd{\deltah}{\bar{\delta}} \newrobustcmd{\sign}{\text{sgn}}

\newrobustcmd{\p}{\mathsf{p}} \newrobustcmd{\g}{\mathsf{g}} \renewrobustcmd{\k}{\mathsf{k}} \newrobustcmd{\K}{\mathcal{K}} \newrobustcmd{\G}{\mathcal{G}} \newrobustcmd{\F}{\mathcal{F}} \newrobustcmd{\E}{\omega} \renewrobustcmd{\S}{\mathcal{S}} 
\renewrobustcmd{\H}{\mathcal{H}}   \newrobustcmd{\T}{\mathcal{T}} 
\newrobustcmd{\one}{\mathds{1}}   
\newrobustcmd{\ones}{\mathcal{I}}

\newrobustcmd{\ket}[1]{|#1\rangle}
\newrobustcmd{\bra}[1]{\langle#1|}
\newrobustcmd{\brkt}[1]{\langle #1 \rangle}
\newrobustcmd{\braket}[2]{\langle #1 | #2 \rangle}

\newrobustcmd{\Ket}[1]{\bm{|}#1\bm{)}}
\newrobustcmd{\Bra}[1]{\bm{(}#1\bm{|}}
\newrobustcmd{\Braket}[2]{\bm{(}#1\bm{|}#2\bm{)}}
\newrobustcmd{\Brkt}[1]{\bm{(} #1 \bm{)}}

\newrobustcmd{\op}[1]{\hat{#1}}
\newrobustcmd{\sop}[1]{\mathcal{#1}}

\DeclareMathOperator{\Tr}{Tr}
\newrobustcmd{\tr}[1]{\underset{#1}{\Tr}}
\newrobustcmd{\tri}[1]{{\Tr}_{#1}}

\newrobustcmd{\Krate}{\lambda_\mathcal{K}}
\newrobustcmd{\Grate}{\lambda_\mathcal{G}}
\newrobustcmd{\Prate}{\lambda_{\Pi}}

\newrobustcmd{\Eq}[1]{Eq.~(\ref{#1})}
\newrobustcmd{\Eqs}[1]{Eqs.~(\ref{#1})}
\newrobustcmd{\eq}[1]{(\ref{#1})}
\newrobustcmd{\Fig}[1]{Fig.~\ref{#1}}
\newrobustcmd{\fig}[1]{\ref{#1}}
\newrobustcmd{\Figs}[1]{Figs.~\ref{#1}}
\newrobustcmd{\Sec}[1]{Sec.~\ref{#1}}
\newrobustcmd{\App}[1]{App.~\ref{#1}}
\newrobustcmd{\app}[1]{\ref{#1}}
\ifdefined\Ref
\renewrobustcmd{\Ref}[1]{Ref.~[\onlinecite{#1}]}
\else
\newrobustcmd{\Ref}[1]{Ref.~[\onlinecite{#1}]}
\fi
\newrobustcmd{\Refs}[1]{Refs.~[\onlinecite{#1}]}

 \definecolor{mygreen}{rgb}{0,0.6,0}
\definecolor{myorange}{rgb}{1, 0.6, 0}

\newcommand{\todo}[1]{}
\newcommand{\refb}[1]{#1}{}

\newcommand{\refd}[1]{{#1}}
\newcommand{\refall}[1]{{#1}}

\usepackage{hyperref} \hypersetup{colorlinks=true,linktoc=all,linkcolor=blue,breaklinks=false,citecolor=blue,urlcolor=blue}

\usepackage{graphicx}

\begin{document}

\title{
	How quantum evolution with memory is generated in a time-local way
}
\author{K. Nestmann$^{(1,2)}$}
\author{V. Bruch$^{(1,2)}$}
\author{M. R. Wegewijs$^{(1,2,3)}$}
\affiliation{
  (1) Institute for Theory of Statistical Physics,
      RWTH Aachen, 52056 Aachen,  Germany
  \\
  (2) JARA-FIT, 52056 Aachen, Germany
  \\
  (3) Peter Gr{\"u}nberg Institut,
  Forschungszentrum J{\"u}lich, 52425 J{\"u}lich,  Germany
}
\pacs{
}

\begin{abstract}
  Two widely used but distinct approaches to the dynamics of open quantum systems
are the  Nakajima-Zwanzig and time-convolutionless quantum master equation, respectively.
Although both describe identical quantum evolutions with strong memory effects,
the first uses a \emph{time-nonlocal} memory kernel $\K$,
whereas the second achieves the same using a \emph{time-local} generator $\G$. 
Here we show that the two are connected by a simple yet general fixed-point relation:
$\G = \hat{\K}[\G]$.
This allows one to extract nontrivial relations between the two completely different ways of computing the time-evolution and combine their strengths.
We first discuss the stationary generator, which enables a Markov approximation that is both nonperturbative and completely positive for a large class of evolutions.
We show that this generator is \emph{not} equal to the low-frequency limit of the memory kernel, but additionally \enquote{samples} it at \emph{nonzero} characteristic frequencies. This clarifies the subtle roles of frequency dependence and semigroup factorization in existing Markov approximation strategies.
Second, we prove that the fixed-point equation sums up the time-domain gradient / Moyal expansion for the time-nonlocal quantum master equation, providing nonperturbative insight into the generation of memory effects.
Finally, we show that the fixed-point relation enables a \emph{direct} iterative numerical computation of both the stationary and the transient generator from a given memory kernel.
For the transient generator this produces non-semigroup approximations which are constrained to be both \emph{initially} and \emph{asymptotically} accurate at each iteration step. \end{abstract}

\maketitle
\section{Introduction\label{sec:intro}}
It is well-known that the dynamics $\rho(t_0) \to \rho(t)$ of the state of an open quantum system
initially uncorrelated with its environment
can be described equivalently by two \emph{exact}, but  fundamentally different quantum master equations (QMEs).
On the one hand, the Nakajima-Zwanzig~\cite{Nakajima58,Zwanzig60} \emph{time-nonlocal} QME
\begin{align}
	\tfrac{d}{dt}\rho(t) = -i \int_{t_0}^t ds \K(t,s) \rho(s)
	\label{eq:qme-nonlocal}
\end{align}
features a \emph{memory kernel} $\K(t,s) $ with separate dependence on all intermediate times $s \in [t_0,t]$. \refall{Here memory is simply understood as retardation.}
On the other hand, the time-convolutionless \emph{time-local} QME of Tokuyama and Mori~\cite{Tokuyama75,Tokuyama76}
\begin{align}
	\tfrac{d}{dt}\rho(t) = -i \G(t,t_0) \rho(t)
	\label{eq:qme-local}
\end{align}
has a \emph{generator} $\G(t,t_0)$, which incorporates the memory integral into its dependence on the current time $t$ and the initial time $t_0$.
Both equations are widely used in the areas of quantum transport,
chemical kinetics,
quantum optics
and quantum-information theory.
In the absence of coupling to the environment and external driving
there is a simple relation between the two:
$\K(t-s)=L \bar\delta(t-s)$ is time-local~\footnote
	{We use the normalization $\int_0^{\infty} ds \bar\delta(s)=1$ for initial-value problems.}
while $\G(t-t_0)=L$ is time independent, such that both reproduce the Liouville-von Neumann equation $\tfrac{d}{dt}\rho(t) = -i [H,\rho(t)]\eqqcolon-iL \rho(t)$ for a closed system.
In some well-understood cases, for example in the limit of weak coupling~\cite{BreuerPetruccione}, high temperature~\cite{Saptsov12,Saptsov14} and limits of singular coupling~\cite{Davies,Gurvitz96,Oguri13}, this simple relation continues to hold, since the Liouvillian $L$ is merely extended by a constant term accounting for dissipative effects,
 $\K(t-s)=\G \bar\delta(t-s)$ with $\G = L+i \mathcal{D}$.
In these cases the time-local QME \eq{eq:qme-local} takes the celebrated Gorini-Kossakowski-Sudarshan~\cite{GKS76}-Lindblad~\cite{Lindblad76} (GKSL) form.
We are interested instead in the generic relation between $\G$ and $\K$ beyond these simple cases,
where strong coupling, low temperature, driving and nonequilibrium non-trivially compete
and both dissipation and memory effects are strong.
Not only are these phenomena important for understanding the disturbance of quantum devices in applications, it is also of intrinsic interest to study them in the highly controlled engineered structures available nowadays~\cite{Barreiro11,Blatt12,Gross17}.

An immediate question is why one would bother to convert between two equivalent QMEs, if instead one could just solve the equation one has in hand for $\rho(t)$?
Careful consideration of this question supports a complementary view~\cite{Vacchini10,Smirne10, Megier17}.
\refall{Typically} $\K$ is easier to compute and
advanced methods have been developed to obtain it analytically~\cite{Schoeller09,Schoeller18} and numerically~\cite{Cohen11,Kidon18} with successful applications to nontrivial models~\cite{Pletyukhov12a,Wilner13,Lindner18,Lindner19}
covering transient and stationary dynamics, as well as counting statistics~\cite{Braggio06,Flindt08, Thomas13} \refall{of observables}. The direct computation of $\G$ using the time-convolutionless formalism~\cite{Shibata77,Shibata80,Chaturvedi79,BreuerPetruccione,Breuer01,Timm11} is typically more challenging.

However, when solving the time-nonlocal equation~\eq{eq:qme-nonlocal}, \refall{taking the frequency-dependence of the memory kernel into account (retardation), may in fact lead one to first construct a corresponding time-local equation~\eq{eq:qme-local}~\cite{Splettstoesser06,Braggio06,Contreras12,Karlewski14}}, which is subsequently solved.
Moreover, the generator $\G$ \emph{by itself} is of particular interest:
it allows to infer important properties of the propagator,
\begin{align}
\rho(t) = \Pi(t,t_0) \rho(t_0)
,
\label{eq:propagator}
\end{align}
which are very difficult to see otherwise.
\refall{
For example, the complete positivity (CP) of the propagator
$\Pi(t,t_0)$, fundamental to its physical legitimacy, may in many
situations beyond the GKSL case be inferred~\cite{Rivas10,Chruscinski12a,Rivas14}
explicitly from a time-dependent canonical form~\cite{Hall14} of $\G$.
This is important for constructing both well defined phenomenological
QMEs~\cite{Vacchini10, Smirne10}
and microscopic models that obey prescribed QMEs~\cite{Amato2019}.
Related to this is that $\G$ often has a clear operational meaning in terms of quantum jumps, which makes it
advantageous for stochastic simulations. For the same reason it is often
employed to construct noise models in quantum-information theory, an
issue of ever increasing importance.
Despite continued efforts, the above is much more complicated to achieve when using $\K$,
either via its microscopic coupling expansion~\cite{Reimer19a}
or a legitimate-pair
decomposition~\cite{Chruscinski16,Chruscinski17b} encompassing broad
classes of models
(semi-Markov~\cite{Vacchini11,Vacchini13},
collision-models~\cite{Ciccarello13} and
beyond~\cite{Wudarski15,Siudzinska17}).
A further key property that can be inferred directly from $\G$ is its so-called divisibility using its canonical jump
rates~\cite{Rivas10,Chruscinski11,Rivas14} and
jump operators~\cite{Wissmann15,Bae2016}.
Again, this seems prohibitively difficult when using $\K$~\cite{Filippov18}.
Divisibility is not only of key importance for the precise
characterization of
quantum non-Markovianity, see \Sec{sec:summary}, a concept much broader~\cite{Li18} than memory understood as retardation.
It also features in quantum coding~\cite{Wolf08a,Muller15} and tomography~\cite{Nielsen20}, key
distribution~\cite{Vasile11}, teleportation~\cite{Laine14},
and work extraction by erasure \cite{Bylicka16}, see \Sec{sec:summary}.
Finally, the time-local nature of equation \eq{eq:qme-local} featuring $\G$ is crucial
to access geometric~\cite{Sarandy05,Sarandy06,Krimer19} and possible
topological~\cite{Li14,Riwar19} phases in open-system evolution with
applications to pumping, full-counting
statistics~\cite{Sinitsyn09}, fluctuation relations~\cite{Riwar20}, entropy production~\cite{Chruscinski15,Popovic18} and quantum
thermodynamics~\cite{Uchiyama14,Goold16,Vinjanampathy16}.
Thus, although in principle
Eqs.~\eqref{eq:qme-nonlocal}-\eqref{eq:qme-local} are obviously equivalent,
there are many reasons for \emph{explicitly} understanding their general
relation, motivating recent work~\cite{Megier20}.}

The relation between $\G$ and $\K$ has already been investigated
for time-translational systems
in the stationary limit $t_0 \to -\infty$.
\Refs{Braggio06,Contreras12,Karlewski14} discussed this using a memory expansion, i.e.,
a gradient / Moyal expansion~\cite{Moyal49,Groenewold46,Rammer86,Onoda06} in the time-domain
applied to the density operator.
Such expansions are well developed~\cite{Sternheimer98,Zachos00}
for Wigner- and Green-functions~\cite{Rammer86,Onoda06}
\refall{and time-dependent density functional theory~\cite{Dittmann18,Dittmann19}}.
The mentioned works indicated that the naive physical intuition,
that the long-time limit of QME~\eqref{eq:qme-local}
is equivalent to the low-frequency approximation to QME~\eqref{eq:qme-nonlocal},
is wrong:
The stationary generator
$\G(\infty)=\lim_{t \to \infty} \G(t)$ does \emph{not} coincide with
the zero-frequency limit $\hat{\K}(0) = \lim_{\E \to 0} \hat{\K}(\E)$ of the Laplace-transformed memory kernel,	
\begin{align}
	\hat{\K}(\E) = \lim_{t_0 \to -\infty } \int_{t_0}^{t} dt \, \K(t-t_0) e^{i\E(t-t_0)}
	\label{eq:khat}
	.
\end{align}
As a result, \enquote{natural} Markovian \refall{semigroup} approximations
set up within approach~\eqref{eq:qme-nonlocal} or~\eq{eq:qme-local}, using the \emph{exact} $\hat{\K}(0)$ or $\G(\infty)$ respectively, turn out to be \emph{distinct}. This difference has proven to be important in perturbative studies beyond weak coupling~\cite{Braggio06,Contreras12,Karlewski14}, and \refall{is} even crucial for measurement backaction~\cite{Hell14, Hell16}. \refall{From these studies the difference between $\hat{\K}(0)$ and $\G(\infty)$ appears to be very complicated.}
This also ties in with the broader~\cite{Li18} discussion of \refall{non-Markovianity, where the interesting connection between divisibility, statistical discrimination~\cite{Breuer09a,Buscemi16,Bae2016} and information flow~\cite{Chruscinski11,Breuer16rev,Smirne13,Rivas14,Wissmann15,Bae2016,Benatti17} continues to develop~\cite{Megier17,Breuer18a}.}

A further important step was provided by the proof in \Ref{Timm11} that
$\hat{\K}(0)$ and $\G(\infty)$, despite their difference, both have the \emph{exact} stationary state as a right zero eigenvector.
However, this work was restricted to master equations for probabilities
and also left unanswered the relation between the full eigenspectra of $\G(\infty)$ and $\hat{\K}(\E)$,
which is one of the results established in the present paper.
Such relations are of interest since these eigenspectra \refall{enter advanced calculations~\cite{Pletyukhov12a,Lindner18,Lindner19} and} provide insight into the time-evolution~\cite{Chruscinski17a}, just as the eigenspectra of Hamiltonians do for the evolution of closed systems.
Similar exact relations among the eigenvectors of the memory kernel $\K$ proved to be very useful for simplifying the complicated calculations for strongly coupled, strongly interacting quantum dots \refall{far out of equilibrium~\cite{Saptsov12,Schulenborg16,Bruch20a}.
}

Thus, it is a pressing question of both fundamental and practical interest
how the time-local generator is related to the time-nonlocal memory kernel for a general finite-dimensional open quantum system.
The central result of this paper, presented in \Sec{sec:fixed}, is that this relation takes the surprisingly simple form of a \emph{functional fixed-point equation}
$\G(t,t_0) = \hat{\K}[\G](t,t_0)$. Importantly, it applies to transient dynamics and allows for arbitrary driving.

In \Sec{sec:stationary} we first \refall{explore} the implications for time-translational systems in the long time limit,
where the stationary generator becomes the fixed point of a simpler \emph{function} of superoperators,
$\G(\infty) = \hat{\K}(\G(\infty))$.
This leads to the key insight that $\G(\infty)$ \enquote{samples} the memory kernel $\hat{\K}(\E)$ at a finite number of frequencies. This \emph{completely defines} $\G(\infty)$ and significantly simplifies the connection between the mentioned distinct Markovian approximations.
The sampled frequencies are shown to be exact time-evolution poles,
well-known from the Laplace resolvent technique~\cite{Schoeller09,Schoeller18,Pletyukhov12a} for solving the time-nonlocal equation~\eq{eq:qme-nonlocal},
an entirely different procedure.
The transformation connecting eigenvectors of $\G(\infty)$ and $\hat\K(\E)$ is found to be related to so-called initial-slip correction procedures~\cite{Geigenmuller83,Haake83,Haake85,Gaspard99,Yu00}.
We show that both the stationary and the transient fixed-point equation are self-consistent expressions for the solution of the memory expansion discussed in \Refs{Braggio06,Contreras12,Karlewski14} 
by explicitly constructing and summing this series.

In \Sec{sec:iteration} we show that the fixed-point \refall{equation} can be turned into \refall{two} separate iterative numerical approaches for obtaining the transient
and the stationary generator,
respectively,
from a given memory kernel.
This provides a new starting point for hybrid approaches in which the results of advanced time-nonlocal calculations~\cite{Cohen11,Kidon18,Pletyukhov12a,Schoeller18}
can be plugged into the time-local formalisms \emph{directly},
bypassing the solution $\Pi(t,t_0)$ that ties Eqs.~\eqref{eq:qme-nonlocal} and \eqref{eq:qme-local} together.
\Ref{Kidon18} numerically addressed the converse problem of extracting $\K$ from an evolution generated by $\G$, which analytically seems to be more complicated.

\refb{Finally in \Sec{sec:example}
we explicitly illustrate the derived relation between $\K$ and $\G$
on two \emph{nonperturbative} examples.
For the exactly solvable dissipative Jaynes-Cummings model~\cite{Garraway97,Mazzola09,Vacchini10,BreuerPetruccione}
we show how it deals with nontrivial singularities of $\G(t,t_0)$ in time.
The fermionic resonant level model~\cite{Reimer19b,Bruch20a} with its richer time-dependent algebraic structure further showcases the nontrivial connection between a time-local and nonlocal description.
}

\refb{
We summarize in \Sec{sec:summary}
and outline how our result may enable progress in various directions.
}
Throughout the paper we set $\hbar=k_\text{B}=1$.

\begin{figure*}[t]
	\includegraphics[width=0.75\linewidth]{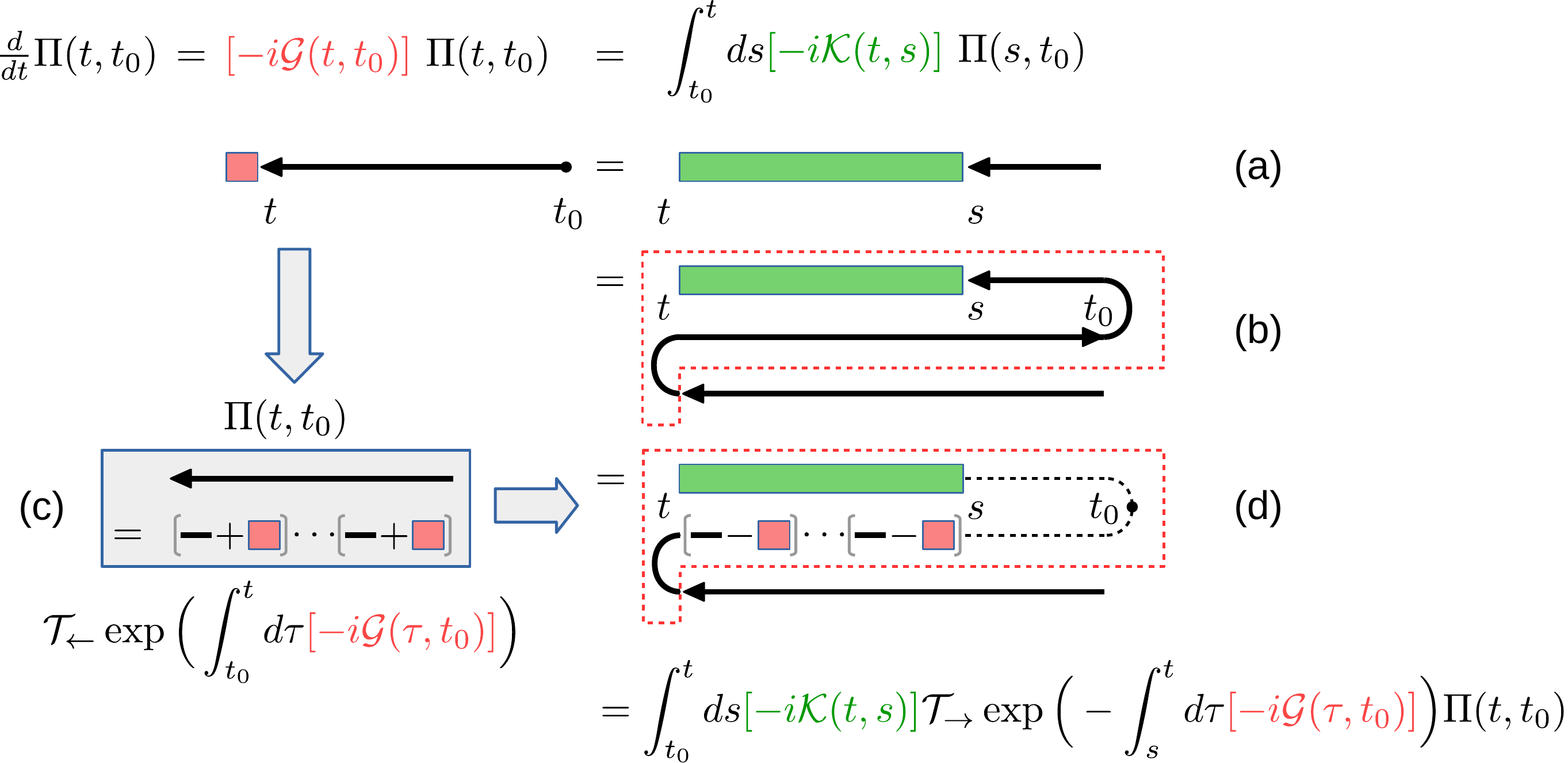}
	\caption{
		Graphical representation of the derivation of the functional fixed-point equation~\eqref{eq:fixed-point}.
		(a)~Equivalent expressions for ${d \Pi(t,t_0)}/{dt}$ as given by the two QMEs.
		(b)~Insertion of canceling backward and forward propagation to \emph{initial time} $t_0$.
		(c)~Evolution
		$\T_{\leftarrow} \exp \big( {\int_{t_0}^{t}d\tau [-i\G(\tau,t_0)]} \big)=
		\lim_{N\to \infty} (\ones -i \G(t_1)\Delta t_1) \ldots (\ones -i \G(t_N)\Delta t_N)$
		expressed as product of infinitesimal steps for the sake of illustration.
		(d)~Backward propagation to memory-time $s$ expressed in terms of $\G$ using the divisor.
		The self-consistency expressed by the functional fixed-point equation~\eqref{eq:fixed-point}
		arises from backward propagation that is needed to enforce the time-local structure of QME~\eqref{eq:qme-local} onto the QME~\eqref{eq:qme-nonlocal}.
		For time-translational systems in the stationary limit the generator becomes $\G(\tau,t_0) \to \G(\infty)$ and literally takes on the role of the complex frequency at which $\hat{\K}(\E)$ is sampled in \Eq{eq:fixed-point-stationary}.
	}
	\label{fig:diagram}
\end{figure*}

\section{Functional fixed-point equation\label{sec:fixed}}

By definition the generator $\G(t,t_0)$ and the memory kernel $\K(t,s)$ are related by the fact that they produce the same dynamics $\Pi(t,t_0)$ [\Eq{eq:propagator}].
To derive a \emph{direct} relation we start from the time-local QME for the propagator,
\begin{align}
\tfrac{d}{dt}\Pi(t,t_0) = -i \G(t,t_0) \Pi(t,t_0),
\label{eq:qme-local-pi}
\end{align}
with initial condition given by identity, $\Pi(t_0,t_0)=\ones$. The generator can be obtained from the above equation
\refall{assuming the inverse propagator exists~\cite{Chruscinski10,Chruscinski18,Chakraborty19,Chakraborty20},}
\begin{align}
-i \G(t,t_0)=  \big[ \tfrac{d}{dt}\Pi(t,t_0) \big] \Pi(t,t_0)^{-1}
\label{eq:generator}
,
\end{align}
\refall{postponing discussion of singular time points to \Sec{sec:jc}.}
The equivalent time-nonlocal QME,
\begin{align}
\tfrac{d}{dt}\Pi(t,t_0) = -i \int_{t_0}^t ds \,  \K(t,s) \Pi(s,t_0)
\label{eq:qme-nonlocal-pi}
,
\end{align}
when inserted into equation~\eq{eq:generator}, gives
\begin{align}
	\G(t,t_0) = \int_{t_0}^t ds \,  \K(t,s) \Pi(s,t_0) \Pi(t,t_0)^{-1}
	\label{eq:generator-rel}
	.
\end{align}
The key step to connect these two approaches originating in statistical physics~\cite{Nakajima58,Zwanzig60,Tokuyama75,Tokuyama76}
is to recognize the expression for the \emph{divisor} $\Pi(t,s|t_0)\coloneqq\Pi(t,t_0) \Pi(s,t_0)^{-1}$.
This quantity is well-known from the quantum-information approach to open-system dynamics, which focuses on complete-positivity (CP) and divisibility properties~\cite{GKS76,Lindblad76,Rivas10,Chruscinski11,Rivas14}.
The divisor describes the propagation  
$\rho(t)= \Pi(t,s|t_0) \rho(s|t_0)$
starting from a state at an \emph{intermediate} time $s \in [t_0,t]$
produced by the same evolution, $\rho(s|t_0)=\Pi(s,t_0)\rho(t_0)$.
This results in the \emph{parametric} dependence on $t_0$.
The divisor obeys the \emph{same} time-local QME, $d\Pi(t,s|t_0)/dt=-i \G(t,t_0) \Pi(t,s|t_0)$, with initial condition $\Pi(s,s|t_0)=\ones$ for all $s \in [t_0,t]$.
The inverse of its formal solution,
\begin{align}
		\Pi(s,t|t_0)
		&=[\Pi(t,s|t_0)]^{-1}
		= \T_{\rightarrow} e^{ i\int^t_s d\tau \G(\tau,t_0)}
		\label{eq:solution-b}
\end{align}
for $t_0 \leq s \leq t$, involves anti-time-ordering denoted by $\T_{\rightarrow}$.
Inserted into \Eq{eq:generator-rel} we find the main result of the paper:
\begin{align}
\G(t,t_0) = \hat{\K}[\G](t,t_0)
.
\label{eq:fixed-point}
\end{align}
The time-local generator is a \emph{fixed point of a functional}
which maps a superoperator function of time $X(t,t_0)$ to another such function:
\begin{align}
\hat{\K}[X](t,t_0) \coloneqq \int_{t_0}^t ds
\,  \K(t,s) 
\T_{\rightarrow} e^{ i\int_s^t d\tau X(\tau,t_0)}
\label{eq:fixed-point-b}
.
\end{align}
This functional is closely related to the ordinary Laplace transform \eq{eq:khat} of the memory kernel $\K(t-s)$, to which it reduces for constant c-number functions of time $X=\E \ones$ in the limit $t-t_0 \rightarrow \infty$ for time translational systems.
We already note that the functional $\hat{\K}[X]$ may have fixed points \emph{other} than $X=\G$.
The nonuniqueness and stability of fixed points are further discussed in \Sec{sec:example} for \refall{two specific models}.

In \Fig{fig:diagram} we graphically outline this derivation. This highlights that time-local propagation with $\G$ needs to be consistent with time-locally evolving backward with $\G$ and \emph{time-nonlocally} propagating forward with the memory kernel.
We stress that \Eq{eq:fixed-point} is a transformation between two complementary descriptions of the same dynamics. It thus also applies to \emph{approximate} dynamics $\Pi'$ generated equivalently by some $\K'$ and $\G'$, and thus has broad applicability.
In the present paper we \refall{aim to highlight the intrinsic functioning of the fixed-point relation and therefore focus on its implications for exactly solvable dynamics.}

Equation~\eq{eq:fixed-point} is explicitly consistent with trace-preservation, a fundamental property of the dynamics.
Due to the ordering in \Eq{eq:fixed-point-b}, where the kernel $\K$ stands to the left of the exponential, the trace-preservation property of the kernel, $\Tr \K(t,s) \bullet =0$, implies the corresponding property of the generator, $\Tr \G(t,t_0)\bullet =0$, where $\bullet$ denotes some operator argument.
In fact, for \emph{any} superoperator function $X(t,t_0)$ one has
\begin{align}
\Tr \hat{\K}[X](t,t_0) = 0
.
\label{eq:iteration-TP}
\end{align}
Moreover, the connection between the  hermicity-preservation property of the kernel and the generator can also be easily checked:
Since $-i \K(t)A=[-i \K(t) A^\dag]^\dag = \H [-i \K(t)] \H A$ for any operator $A$,
where $\H A \coloneqq A^\dag$ is an antilinear superoperator, we have
\begin{align}
	\H \big\{ -i \hat{\K}[X](t,t_0) \big\} \H &
	 = -i \hat{\K}[ -\H X \H ](t,t_0)
	.
	\label{eq:iteration-HP}
\end{align}
 \section{Stationary fixed-point equation\label{sec:stationary}}

We now focus on the implications for time-translational systems in the stationary limit
and consider the case where the generator converges to a constant superoperator
$\G(\infty)= \lim_{t_0 \to - \infty} \G(t-t_0)$.
Then the idea is that at large $t-t_0$
we can replace~\footnote{\refall{The limit $t-t_0 \rightarrow \infty$ converges for the examples in \Sec{sec:example}, but in general requires care and is beyond the present scope.}} the time-ordered exponential in \eqref{eq:fixed-point-b} by an exponential function:
\begin{subequations}
	\begin{align}
	\hat{\K}[\G](t-t_0) & = \int_{t_0}^t ds
	\,  \K(t-s) 
	\T_{\rightarrow} e^{ i\int_s^t d\tau \G(\tau-t_0)}
	\label{eq:step1}\\
	&
	\approx \int_{-\infty}^t ds
	\,  \K(t-s)  e^{ i (t-s) \G(\infty)}.
	\end{align}
	\label{eq:steps}\end{subequations}Here we use that typically either the generator has already become stationary,
$\G(\tau-t_0) \approx \G(\infty)$ ($\tau \geq s \gg t_0$),
or the memory kernel has already decayed ($t \gg s$), thus suppressing the expression.
Hence we obtain the \emph{stationary} fixed-point equation
\begin{align}
	\G(\infty)& =  \hat{\K}(\G(\infty)).
	\label{eq:fixed-point-stationary}
\end{align}
It features instead of \Eq{eq:fixed-point-b} the much simpler extension of the Laplace transform \eq{eq:khat} with frequency $\E$ replaced by the time-constant superoperator $X$:
\begin{align}
\hat{\K}(X)
& =
 \int_{0}^{\infty} ds \,  \K(s) e^{ i s X }
.
\label{eq:fixed-point-stationary-b}\end{align}

\subsection{Exact {sampling} relation between spectral decompositions}
The stationary fixed-point equation \eqref{eq:fixed-point-stationary}
immediately makes clear that in general the stationary generator $\G(\infty)$ is \emph{not} the low-frequency limit of the memory kernel,
$\hat \K(0) = \lim_{\E \to i0^{+} } \hat{\K}(\E)$.
We now make precise \emph{which} parts of the frequency dependence of
the memory kernel $\hat{\K}(\E)$ matter in the stationary limit.
To this end, assume that one can diagonalize
the stationary generator
$
\G(\infty) = \sum_i g_i \Ket{g_i} \Bra{\bar{g}_i},
$
and denote the distinct left and right
eigenvectors to the same eigenvalue $g_i$ by $\Bra{\bar{g}_i}$ and $\Ket{g_i}$ respectively, which
satisfy the Hilbert-Schmidt~\footnote
	{Note that $\Ket{g_i}=\hat{g}_i$ corresponds to an operator $\hat{g}_i$ and that $\Bra{\bar{g}_i}\bullet=\Tr [ \hat{\bar{g}}_i^\dag \bullet]$ corresponds to a \emph{different} operator $\hat{\bar{g}}_i$ -- indicated by the bar -- such that $\Braket{\bar{g}_i}{g_{j}}=\Tr [ \hat{\bar{g}}_i^\dag \hat{g}_{j} ]=\delta_{ij}$.}	
biorthogonality relation $\Braket{\bar{g}_i}{g_{i'}}=\delta_{ii'}$.
Insertion into \Eq{eq:fixed-point-stationary} gives $\G(\infty)=\sum_i \hat{\K}(g_i)  \Ket{g_i} \Bra{\bar{g}_i}$
with the ordinary Laplace transform \eq{eq:khat} evaluated at $\E=g_i$.
Focusing on nondegenerate eigenvalues we therefore have
\begin{align}
	\hat{\K}(g_i)  \Ket{g_i} =\G(\infty) \Ket{g_i} = g_i \Ket{g_i}
	\label{eq:eigenvectors}
	.
\end{align}Diagonalizing the kernel \emph{after} Laplace transforming,
$\hat{\K}(\E) = \sum_j k_j(\E) \Ket{k_j(\E)} \Bra{\bar{k}_j(\E)}$, this implies that
at designated frequencies $\E = g_i$ \emph{one} of its eigenvalues, labeled $j=f_i$,
must coincide with an eigenvalue $g_i$ of the stationary generator $\G(\infty)$:
\begin{align}
k_{f_i}(g_i) = g_i
\label{eq:pole}
.
\end{align}
The right eigenvectors can then be normalized to coincide
\begin{align}
\Ket{k_{f_i}(g_i)} = \Ket{g_i}
.
\label{eq:eigenvector}
\end{align}
\refall{Importantly the eigenvectors of the kernel $\Ket{k_j(\E)}$ can also contain poles, which have an important impact on the evolution as illustrated explicitly in \Sec{sec:rlm}. However since $\G(\infty)$ was assumed to be finite it can not sample any  of these eigen\emph{vector} poles of the kernel.}

We note that the left eigenvectors $\Bra{\bar{g}_i}$ and $\Bra{\bar{k}_{f_i}(g_i)}$ in general \emph{differ}
with one important exception, labeled by $i=0$:
From the trace-preservation property of the dynamics [see \Eq{eq:iteration-TP}] it follows that both $\G(\infty)$ and $\hat{\K}(\E)$ at every frequency $\E$ have the \emph{left} zero eigenvector $\Bra{\one}=\Tr \bullet$, the trace functional. The corresponding zero eigenvalue is denoted by $g_0=k_0(\E)=0$ for all $\E$ labeling $f_0=0$. Thus, a nontrivial consequence of \Eq{eq:eigenvectors} is that the associated \emph{right} zero eigenvectors
of $\G(\infty)$ and $\hat{\K}(0)$, respectively, coincide with the stationary state:
\begin{align}
	\Ket{g_0}=\Ket{k_0(0)}=\Ket{\rho(\infty)}
	\label{eq:eigenvector-stationary}
	.
\end{align}
This generalizes the result of \Ref{Timm11}, which proved this statement for probability vectors evolving with a time-local master equation \refall{(i.e. for probabilities only)}.

We summarize the key result of this section: For Hilbert-space dimension $d$ the stationary time-local generator, with its finite set of eigenvalues $g_0,\ldots,g_{d^2-1}$, can be written as
\begin{subequations}
	\begin{align}
\G(\infty)
= &\sum_{i} k_{f_i}(g_i)  \Ket{k_{f_i}(g_i)} \Bra{\bar{g}_i}.
\label{eq:G-sampling}
\end{align}
It \enquote{samples} \emph{one} term of the Laplace-transformed memory kernel at \emph{each} of the frequencies $\E = g_0,\ldots,g_{d^2-1}$:
\begin{align}
\hat{\K}(g_i)
\notag
&
=
k_{f_i}(g_i)  \Ket{k_{f_i}(g_i)} \Bra{\bar{k}_{f_i}(g_i)}
\\
&
+ \sum_{j\neq {f_i}}
k_j(g_i)  \Ket{k_j(g_i)} \Bra{\bar{k}_j(g_i)}
.
\label{eq:K-sampling}
\end{align}\label{eq:sampling}\end{subequations}From each sampled frequency only a single right eigenvector $\Ket{k_{f_i}(g_i)}$ for one specific eigenvalue satisfying $k_{f_i}(g_i)=g_i$ is needed to construct $\G(\infty)$.
Importantly, its left eigenvectors $\Bra{\bar{g}_i}$ are determined by the right ones through the biorthogonality constraint.

Anticipating later discussion we note that some intuitive ideas turn out to be incorrect:
First, the sampling formula shows that in general \emph{nonzero} frequencies of $\hat \K(\E)$ may matter at stationarity.
It thus makes precise that \enquote{memory}, often understood as \refall{retardation/}\emph{frequency dependence} of the kernel~\cite{Braggio06, Contreras12, Hell14, Karlewski14, Hell16}, is in general \emph{not} the same as \enquote{memory}  defined by a \emph{Markovian semigroup}~\cite{GKS76, Lindblad76, Rivas10, Rivas14, Chruscinski11}, in which $\G(\infty)$ naturally appears as we discuss later.
Second, the sampled frequencies $g_i$ need \emph{not} be the eigenvalues with the smallest decay rates \refall{[$-\Im k_j(\E_p)$], as illustrated in \Sec{sec:example}.
}

\refb{The sampling formula~\eq{eq:sampling}
implies that the analytical calculation of the typically more complicated quantity $\G(\infty)$ can in principle be reduced to the calculation of $\hat{\K}(\E)$ at just $d^2$ specific frequencies. We will show in \Sec{sec:stationary-iteration} how $\G(\infty)$ can be iteratively computed from $\hat{\K}(\E)$, thus determining which frequencies are actually sampled. It is therefore not necessary to compute the transient generator $\G(t)$ in order to compute $\G(\infty)$.}
This is a significant advance since $\hat\K$ can be \refb{approximated accurately} for complicated many-body dynamics using well-developed techniques~\cite{Schoeller09,Pletyukhov12a,Schoeller18,Lindner18,Lindner19}.
\refb{As mentioned, our relations remain valid when dealing with such approximate kernels: they are simply a way to change from a time-nonlocal to a time-local representation.
}

\subsection{Exact time-evolution poles}

We now compare the sampling relation \eq{eq:sampling} with the formal exact solution for time-translational systems obtained by the resolvent method:
Laplace transforming the time-nonlocal QME~\eq{eq:qme-nonlocal}
to obtain the \enquote{Green's function} or resolvent $\hat{\Pi}(\E)={i}/(\E-\hat{\K}(\E))$, and transforming back by integration along a clockwise oriented contour $\mathcal{C}$ closed in the lower half of the complex plane, we get:
\begin{gather}
\Pi(t-t_0)
= \int_{\mathcal{C}} \frac{d\E}{2\pi} \hat{\Pi}(\E) e^{-i \E(t-t_0)}
\label{eq:inverse-laplace}
\\= -i \sum_p \text{Res}\big[\hat{\Pi}(\E_p) e^{-i\E_p(t-t_0)  } \big]
+
\int_{\text{bc}} \frac{d\E}{2\pi} \hat{\Pi}(\E) e^{-i \E(t-t_0)}
\notag
.
\end{gather}
Here $\text{Res}\big[f(\E_p) \big]$ is the residue at pole $\E_p$
and ``b.c.'' indicates integration over possible branch cut contributions of $\hat{\Pi}(\E)$,
see \Refs{Schoeller09,Schoeller18,Andergassen11a,Pletyukhov12a} for details and applications.
The eigenvalue poles of $\hat{\Pi}(\E)$ solve the equation $\E_p=k_j(\E_p)$ for some eigenvalue of $\hat{\K}$.
By our result \eq{eq:pole} the eigenvalues of $\G(\infty)$ are guaranteed to be included among these \emph{eigenvalue} poles of $\hat\Pi(\E)$.
Thus, our stationary fixed-point equation~\eq{eq:fixed-point-stationary} reveals how the time-local approach keeps track of these characteristic frequencies of the evolution, which are explicit in the time-nonlocal approach.

In other words, for time-translational systems
the relation
$\G(\infty) = \hat{\K}(\G(\infty))$
establishes that the time-local generator $\G(\infty)$ is a superoperator-valued characteristic \enquote{frequency} of the evolution.
To be sure, there are further contributions from non-sampled poles and branch cuts, which can be infinitely many and may also involve the eigenvectors~\cite{Schoeller18}. These are encoded in the transient fixed-point \refall{equation~\eq{eq:fixed-point} through the anti-time-ordered integration \eq{eq:fixed-point-b}.}
Thus, the eigenvalues of $\G(\infty)$ generally do not exhaust all the \refall{eigenvalue} poles of $\hat{\Pi}(\E)$.
\refall{\emph{Which} of the eigenvalues of $\hat \K(\E)$ satisfying $\E_p = k_j(\E_p)$ are
eigenvalues of $\G(\infty)$ is not \emph{apriori} clear, as discussed above}.

Our result \eq{eq:G-sampling} now reveals that
the first contribution to the exact dynamics~\eq{eq:inverse-laplace}
actually contains a Markovian semigroup exponential:
\begin{gather}
 \Pi(t-t_0)
= e^{-i(t-t_0) \G(\infty)} \S + \ldots
\label{eq:inverse-laplace2}
\end{gather}
where $\ldots$ denotes the above mentioned non-sampled contributions.
If $\G(\infty)$ exists,
one might expect that the evolution for long times
will eventually follow this semigroup dynamics.
However, this exponential term is already modified by
the time-constant superoperator
\begin{align}
\S =
\sum_{i} \frac{1}{1-\frac{\partial k_{f_i}}{\partial \E}(g_i)}
\Ket{g_i} \Bra{\bar{k}_{{f_i}} (g_i)}
\label{eq:S}
\end{align}
obtained from the residues~\footnote{
Here we are assuming for simplicity that all poles of $\hat\Pi(\omega)$ are first order poles. For a detailed discussion see~\Ref{Bruch20a}.
} in \Eq{eq:inverse-laplace} using \Eq{eq:eigenvector}.
The superoperator $\S$ is of practical importance as it relates to the so-called slippage of the initial condition, a well-known procedure for improving Markovian approximations~\cite{Geigenmuller83,Haake83,Haake85,Gaspard99,Yu00}, \refall{see discussion in \Sec{sec:summary}}.

\subsection{Nonperturbative semigroup approximations\label{sec:approx}}

We can now address the puzzling issue regarding the more basic approximation strategy that we mentioned in the introduction:
The \emph{equivalent} QMEs~\eqref{eq:qme-nonlocal} and \eqref{eq:qme-local} \enquote{naturally} lead to semigroup approximations which \emph{differ}, even when constructed from the \emph{exact} $\G$ and $\K$.

(i) \emph{Stationary generator} $\G(\infty)$:
Assuming that the generator converges to a stationary value $\G(\infty)$
we can try to approximate the time-local QME~\eq{eq:qme-local} for large $t$
by replacing the generator by its constant stationary value,
$
\tfrac{d}{dt}\rho(t) \approx -i \G(\infty) \rho(t)
$.
This idea underlies~\Refs{Contreras12,Karlewski14}
and motivated the direct calculation of $\G(\infty)$ by a series expansion in the coupling in \Ref{Timm11}.
The resulting approximate dynamics
\begin{subequations}
	\begin{align}
	\Pi(t,t_0) & \approx e^{-i(t-t_0)\G(\infty)}
	\\
	& = \sum_i e^{-i g_i(t-t_0)} \Ket{g_i} \Bra{\bar{g}_i}
	\label{eq:approx-G-b}
	.
	\end{align}\label{eq:approx-G}\end{subequations}has an interesting feature: There are many evolutions for which the \emph{asymptotic} generator  $\G(\infty)$ has a GKSL form~\cite{GKS76,Lindblad76} with nonnegative \refall{jump rates},
which guarantees that the approximation is completely positive in addition to trace preserving.
Nonperturbative approximations preserving \emph{both} these properties are notoriously difficult to construct, especially starting from microscopic models~\cite{vanWonderen13,vanWonderen18a,vanWonderen18b,Reimer19a}.
Here the class of evolutions goes beyond semigroups
by including all CP-divisible evolutions, but also allowing for certain \emph{non} CP-divisible ones~\footnote
	{Note that $\Pi(t)$ is CP-divisible~\cite{Rivas10,Chruscinski11,Rivas14} if $\G(t)$ has nonnegative GKSL coefficients for \emph{all} times $t$. We allow for negative coefficients at \emph{finite} times. Models where the evolution has negative asymptotic GKSL coefficients have also been studied recently~\cite{Hall14,Megier17,Siudzinska20}}.

Our sampling result~\eq{eq:sampling} allows this to be compared with a corresponding approximation in the Laplace resolvent approach to the \emph{time-nonlocal} QME~\eq{eq:qme-nonlocal}:
if one keeps the first term of \Eq{eq:inverse-laplace} and selects only the fixed-point poles $\E=g_i$,
then one obtains the semigroup approximation together with the initial-slip correction $\S$ as in \Eq{eq:inverse-laplace2}.
Due to the automatic inclusion of $\S$, this approximation is neither a semigroup
nor a CP map around the initial time $t_0$~\cite{Bruch20a}.
This may give faster convergence but also fail dramatically\refall{, see discussion in \Sec{sec:summary}}.
In contrast, the semigroup approximation \eq{eq:approx-G} does not suffer from such problems.

(ii) \emph{Low-frequency memory kernel $\hat{\K}(0)$}.
Starting instead from the time-nonlocal QME~\eq{eq:qme-nonlocal}, one may argue that for slowly varying dynamics only the low-frequency part of the memory kernel matters. Replacing $\rho(s) \to \rho(t)$ in the integrand and taking $t_0 \to -\infty$, one then obtains
$	\dot{\rho}(t)
	\approx
-i \hat{\K}(0) \rho(t)
$
with the approximate solution
\begin{subequations}
	\begin{align}
	\Pi(t,t_0) & \approx e^{-i (t-t_0) \hat{\K}(0)}
	\label{eq:approx-K-a}
	\\&=
	\sum_j e^{-i k_j(0)(t-t_0)} \Ket{k_j(0)} \Bra{\bar{k}_j(0)}.
\end{align}\label{eq:approx-K}\end{subequations}In the resolvent approach this approximation is equivalent to neglecting all frequency dependence of the memory kernel,
$\hat{\Pi}(\E) \approx i/(\E - \hat{\K}(0))$,
leaving only $d^2$ eigenvalue poles $\E_j = k_j(0)$.
In contrast to $\G(\infty)$,
we know of no general conditions that guarantee that $\hat{\K}(0)$ generates a completely positive evolution for some broad class of nontrivial models.
Even when it is known that $\G(\infty)$ has nonnegative GKSL coefficients -- ensuring \Eq{eq:approx-G} is completely positive --
one \emph{still} has to explicitly check that the same holds for $\hat{\K}(0)$.
\refall{Although} both approximations \eq{eq:approx-G} and \eqref{eq:approx-K} nonperturbatively account for oscillation frequencies and decay rates \refall{in a different way}, it follows from the sampling result~\eq{eq:eigenvector} that \emph{both} converge to the \emph{exact} stationary state.
In \Sec{sec:example} we will illustrate their difference. Note, however, that $\G(\infty)=\hat{\K}(0)$ is possible also for
\refall{a non-semigroup}
evolution
[\Eq{eq:trivial}].

\subsection{Summing the memory expansion\label{sec:summing}}

Whereas the argument leading to \Eq{eq:approx-K} may be justified in the weak coupling limit,
it has been noted that when computing $\hat{\K}$ to higher order in the system-environment coupling this becomes inconsistent~\cite{Splettstoesser06,Cavaliere09,Contreras12,Karlewski14}.
In terms of \Eq{eq:generator-rel} this means that one must not only expand the kernel $\K(t-s)$  in the memory-time $s$ relative to the current time $t$, but simultaneously expand $\Pi(s,t_0)=\Pi(t,t_0)-(t-s)\partial\Pi(t,t_0)/\partial t + \ldots $ under the memory integral.
This way \Ref{Contreras12} obtained a stationary time-local QME with an approximate generator
\begin{align}
	\G(\infty) \approx \hat{\K}(0) +  \frac{\partial \hat{\K}}{\partial \E}(0) \,  \hat{\K}(0).
	\label{eq:adiabatic}
\end{align}
When computing $\hat{\K}(0)$ to second order in, e.g., a tunnel coupling, the first order contributions to the second term are comparable~\cite{Contreras12} and may lead to cancellations that are necessary to respect complete positivity~\cite{Hell14,Hell16}.

One may roughly understand \Eq{eq:adiabatic} as follows: to obtain $\G(\infty)$ one linearizes the frequency dependence of the memory kernel
$\hat{\K}(\E) \approx \hat{\K}(0) +[ {\partial \hat{\K}}/{\partial \E}(0) ]  \E$ and evaluates it
at the characteristic \enquote{frequency} $\E=\G(\infty) \approx \hat{\K}(0)$ of the system, which in first approximation is the low-frequency kernel itself.
This tentative picture is made rigorous by our fixed-point equation \eqref{eq:fixed-point-stationary},
where the frequency is likewise replaced by a superoperator, but in a self-consistent way.
In \Ref{Karlewski14} the approximation \eq{eq:adiabatic} was generalized to higher orders by applying partial integrations of the time-nonlocal QME~\eqref{eq:qme-local}, which can be shown to be equivalent to further continuing the memory expansion of \Ref{Contreras12}.
\refall{In App. C we show how this gradient expansion can be expressed in Moyal brack-
ets~\cite{Moyal49,Groenewold46} with respect to time
similar to that used in Green’s function
techniques~\cite{Rammer86,Onoda06}.
It has also recently been used to combine QMEs with time-dependent density-functional theory~\cite{Dittmann18,Dittmann19}.
}

Thus, starting from the time-nonlocal QME
one is led to a \emph{time-local} QME by a memory expansion~\eq{eq:approx-K}.
Another key result of this paper is that this series can in fact be summed up to all orders as we show in~\App{app:memory}.
One finds that the constant generator that accounts for 
\emph{all} memory terms of the stationary \emph{time-nonlocal} QME is the stationary time-local generator obeying $\G(\infty)=\hat{\K}(\G(\infty))$, our stationary fixed-point equation~\eq{eq:fixed-point-stationary}.
This means that our sampling formula~\eq{eq:sampling} is the nonperturbative result of this memory expansion:
The infinite sum of memory terms -- featuring all derivatives of $\hat{\K}(\E)$ at \emph{zero} frequency -- can be condensed into a finite sum of contributions of $\hat{\K}(\E)$ at just $d^2$ finite frequencies $\E=g_i$.

Importantly, the memory expansion can even be summed up for the full transient dynamics,
thereby recovering $\G(t,t_0) = \hat{\K}[\G](t,t_0)$, the functional fixed point equation~\eq{eq:fixed-point} [\App{app:memory}].
By making use of the divisor we can give a closed formula for terms of arbitrary order [\Eqs{eq:Fk},~\eq{eq:F-coef}].
Altogether, this shows that equations~\eq{eq:fixed-point} and \eq{eq:fixed-point-stationary} are very useful for generating gradient expansions in time when given a memory kernel $\K$.
We next explore a different approach where one solves for the transient $\G(t-t_0)$, giving approximate evolutions which are not semigroups as \Eq{eq:approx-G} and \eq{eq:approx-K}.
\refall{Also there} \emph{both} $\G(\infty)$ and $\hat \K(0)$ play an interesting role.
 \section{Iterative construction of generator from memory kernel\label{sec:iteration}}

Our final key result is that the fixed-point equation may be turned into a computational tool
to obtain $\G$ from a given memory kernel $\K$ computed using a method of choice.
We focus on time-translational systems
-- setting $t_0 =0$ --
and the ideal situation where $\K$ has been computed exactly.

\subsection{Iteration for stationary generator\label{sec:stationary-iteration}}

The simplest scenario is where one iteratively solves~\Eq{eq:fixed-point-stationary}
to find $\G(\infty)$ \emph{directly} from $\K(t)$ or $\hat{\K}(\E)$,
i.e., without considering the transient evolution $\Pi(t)$ or the transient generator $\G(t)$.
Using the converged result one may then set up the nonperturbative semigroup \eq{eq:approx-G} to approximate the full evolution $\Pi(t)$.

First, consider the low-frequency kernel as an initial approximation to the generator,
$\G^{(0)}(\infty) = \hat{\K}(0)$, as in \Eq{eq:adiabatic}.
If the exact dynamics is a semigroup, $\K(t)= \hat{\K}(0) \deltah(t)$
and $\G(t)=\hat{\K}(0)$,
then this already is the fixed point
since $\G^{(1)}(\infty)=\hat{\K}(\hat{\K}(0))=\hat{\K}(0)$.
This may \emph{also} happen for non-semigroup evolutions [\Eq{eq:G-simple}].
In general, further approximations are obtained by $n$-fold iteration,
$\G^{(n)}(\infty) = \hat{\K}\big(\ldots \hat{\K}( \hat{\K}(0) ) \big)$.
Inspecting the first iteration,
\begin{align}
		& \G^{(1)}(\infty) = \hat{\K}(\hat{\K}(0)) = \sum_{j\neq 0} \hat{\K}(k_j(0)) \Ket{k_j(0)}\Bra{\bar{k}_j(0)}
		,
\end{align}
we see that the stationary state $\Ket{k_0(0)}$ remains unaffected  (trace-preservation), 
but in general all $j\neq 0$ contributions are altered by the memory kernel evaluated at \emph{finite} frequencies,
thus generating a difference between $\hat\K(0)$ and $\G(\infty)$.

The convergence of this procedure with $n$ is certainly not obvious, 
but our first applications in \Sec{sec:example} are encouraging.
Indeed, one can consider starting the iteration from \emph{any} initial superoperator $\G^{(0)}(\infty)=X$.
In this case, property~\eq{eq:iteration-TP} guarantees that the iteration trajectory $\G^{(n)}(\infty) = \hat{\K}\big(\ldots \hat{\K}( X ) \big)$ is confined to the linear space of trace-preserving superoperators  irrespective of $X$.
If $iX$ is hermicity-preserving, then the trajectory will additionally be confined to such superoperators by property \eq{eq:iteration-HP}.

\subsection{Functional iteration for transient generator\label{sec:time_dependent_iteration}}

We next describe the more complicated iteration of the functional equation~\eq{eq:fixed-point}.
Here the aim is to construct the full transient generator $\G(t)$ starting from the memory kernel $\K(t)$.
As a preparation we decompose the kernel into its time-local ($\bar{\delta}$-singular) part
and a remaining time-nonlocal part:
\begin{align}
	\K(t) = \K_\l \bar\delta(t) + \K_\n(t).
	\label{eq:Kdecom}
\end{align}
In addition to the system Liouvillian $L$, the part $\K_\l$ may contain an environment-induced contribution (as for fermionic wide-band models~\cite{Saptsov12,Saptsov14,Schulenborg16} as studied in \Sec{sec:rlm}),
but this need not be the case (as in the model studied in \Sec{sec:jc}).
Inserting \Eq{eq:Kdecom} into the functional \eqref{eq:fixed-point-b}
we obtain
\begin{align}
	\G^{(n+1)}(t) = \K_\l +
	\int_{0}^t ds
	\,  \K_\n(t-s) 
	\T_{\rightarrow} e^{ i\int_s^t d\tau \G^{(n)}(\tau)}.
	\label{eq:fixed_point_decomposed}
\end{align}
Iterating this equation starting from the constant function $\G^{(0)}(t) =\hat{\K}(0)$ gives approximations $\G^{(n)}(t)$ which generate evolutions with two important properties \emph{at every iteration}:

First, each approximation is accurate at long times, provided $\G(t)$ has a stationary limit and \Eq{eq:fixed-point} converges to \Eq{eq:fixed-point-stationary}.
Our choice of starting point ensures by \Eq{eq:eigenvector-stationary} that 
$ \G^{(n)}(t)\Ket{\rho(\infty)} =0$ holds initially for $n=0$,
implying that the generated evolution goes to the exact stationary state \refall{for $t \rightarrow \infty$}.
Arguing as in \Eq{eq:steps} we find that this also holds for the next iteration:
$ \G^{(n+1)}(\infty)\Ket{\rho(\infty)}
=
\lim_{t\to \infty} [ \K_\l + \int_{0}^t ds \,  \K_\n(t-s) ] \Ket{\rho(\infty)}
=  \hat{\K}(0)  \Ket{\rho(\infty)} =0
$.
The same argument also applies for starting point $\G^{(0)}(t)=\G(\infty)$ [\Eq{eq:eigenvector-stationary}] or any starting point $X$ for which $X \Ket{\rho(\infty)}=0$. However, starting from the memory kernel formalism, $\hat \K(0)$ is already available.

Second, each generated approximation is also accurate at short times.
To see this, note that at the initial time the generator is given by the time-local part of the kernel
\begin{align}
\G(0) = \K_\l
	,
	\label{eq:G-initial}
\end{align}
which we split off from the generator,
\begin{align}
\G(t) = \K_\l + \G_\n(t)
,
\qquad
\G_\n(0) = 0
.
\label{eq:Gsplit}
\end{align}The second term incorporates all effects due to the time-nonlocal part of the kernel $\K_\n(t)$.
For the first iteration we have
\begin{subequations}
	\begin{align}
\G^{(1)}(t) &= \K_\l  + \int_{0}^{t} ds \, \K_\n(t-s) e^{i \hat{\K}(0)  (t-s)}
\label{eq:here}
\\
& 
\approx \K_\l  + t \, \K_\n(0) + \ldots
\label{eq:explicit_first_iteration}
\end{align}
\end{subequations}
as dictated by the short-time limit of the \emph{time-nonlocal} part of the memory kernel. This implies that in the exponential of the next iteration
we similarly have at short times $\int_s^t d\tau \G^{(1)}(\tau) \approx (t-s) \K_\l $, giving the same leading behavior.
Thus, each iteration $n \geq 1$ coincides with the exact initial generator~\eq{eq:G-initial}
including the \emph{linear} order,
$\G^{(n)}(t) = \K_\l  + t \, \K_\n(0) + \ldots $.
Clearly, no semigroup approximation can achieve this.

\refall{The convergence of this iteration is again not evident and
an analysis of the local stability is complicated due to the time-nonlocality of the superoperator equations.
Remarkably, we numerically find for several models that this procedure can be made to work, even when the generator is time-singular [\Sec{sec:jc}] or has time-dependent algebraic structure [\Sec{sec:rlm}].}
 \section{Examples \label{sec:example}}

\refb{
\subsection{Dissipative Jaynes-Cummings model\label{sec:jc}}

We first illustrate our findings for the} dissipative Jaynes-Cummings model~\cite{Garraway97,BreuerPetruccione,Mazzola09,Vacchini10},
\refb{which is algebraically simple but can show challenging time-singularities in the generator.
}This exactly solvable model describes a two-level atom with transition frequency $\varepsilon$ ($H=\varepsilon d^\dag d$ with $\{d,d^\dag\}=\one$)
interacting with a continuous bosonic reservoir ($H_\text{R} = \int d \omega \omega b_\omega^\dag b_\omega$ with $[b_\omega,b_{\omega'}^\dag]= \delta(\omega-\omega') \one$)
initially in a vacuum state $\ket{0}$.
The coupling is bilinear,
\begin{align}
	H_\text{T} = \int d\omega \sqrt{\frac{\Gamma(\omega)}{2\pi}} \Big( d^\dagger b_\omega + b^\dagger_\omega d \Big)
	\label{eq:coupling}
	,
\end{align}
with real amplitudes set by a spectral density $\Gamma(\omega)$.
The occupation numbers of reservoir modes are either 0 or 1 due to a dynamical constraint:
the coupling~\eq{eq:coupling} conserves the total
excitation number
$d^\dag d + \int d \omega b_\omega^\dag b_\omega$.
\refb{Here we} study the effects of energy-dependent coupling $\Gamma(\omega)$ without initial reservoir statistics ($T=0$):
We assume a Lorentzian profile of width $\gamma$
whose maximum value $\Gamma \equiv \Gamma(\varepsilon)$
lies precisely at the atomic resonance:
\begin{align}
\Gamma(\omega) = \Gamma \frac{ \gamma^2}{(\varepsilon-\omega)^2 + \gamma^2}
.
\label{eq:jc_model_lorentzian}
\end{align}
Although this model has been studied in detail~\cite{Garraway97,Mazzola09,Smirne10,Vacchini10} and features in text books~\cite{BreuerPetruccione}
the remarkable relation between its generator $\G$ and memory kernel $\K$ has not been noted\refb{, but see \Ref{Megier20}}.
All results below can be generalized to any profile $\Gamma(\E)$.

From the solution~\cite{BreuerPetruccione} of the total-system state $\ket{\psi_\text{tot}(t)}$,
with $\ket{\psi_\text{tot}(0)}=\ket{\psi(0)} \otimes \ket{0}$, we extract the propagator
$\Tr_\text{R}\{\ket{\psi_\text{tot}(t)}\bra{\psi_\text{tot}(t)}\}=\Pi(t)\ket{\psi(0)}\bra{\psi(0)}$
working in the Schr\"odinger picture and setting $t_0=0$.
It has the form of an amplitude damping channel~\cite{NielsenChuang}
with spectral decomposition
\begin{align}\Pi(t) = 
	&
	\Ket{00} \Big[ \Bra{00} + \Bra{11} \Big]
	+ |\pi(t)|^2 \Big[\Ket{11}-\Ket{00}\Big] \Bra{11}
	\notag
	\\
	&+ \pi(t) \,  \Ket{01}\Bra{01}
	 + \pi(t)^* \Ket{10}\Bra{10}
	,
	\label{eq:jc_model_pi}\end{align}using $\Ket{\nu\nu'} = \ket{\nu}\bra{\nu'}$ and $\Bra{\nu\nu'}=\bra{\nu}\bullet\ket{\nu'}$,
where
$\ket{\nu}$ denotes the atomic state $\nu=0,1$.
The time-dependent parameter reads
\begin{align}
	\pi(t) \equiv e^{-i\varepsilon t} e^{-\gamma t/2} \left[ \cosh \left(\frac{\gamma' t}{2} \right) + \frac{\gamma}{\gamma'} \sinh\left(\frac{\gamma' t}{2}\right) \right]
	\label{eq:jc_model_g}
\end{align}where $\gamma' \coloneqq \sqrt{\gamma(\gamma -2\Gamma)}$.
Thus, an initially excited state evolves with probability $\bra{1}\rho(t)\ket{1}=|\pi(t)|^2$.
In the frequency domain we have
\begin{align}\hat\Pi(\E) = 
	&
	 \frac{i}{\E}\Ket{00} \Big[ \Bra{00} + \Bra{11} \Big]
	+ \widehat{|\pi|^2}(\E) \Big[\Ket{11}-\Ket{00}\Big] \Bra{11}
	\notag
	\\
	&+ \widehat{\pi}(\E) \Ket{01}\Bra{01} + \widehat{\pi^*}(\E) \Ket{10}\Bra{10}
	.
	\label{eq:pi-frequency}\end{align}The Laplace transforms
\begin{subequations}\begin{align}\widehat{|\pi|^2}(\E) &=
	\frac{1}{4}\frac{(\gamma/\gamma'-1)^2}{\gamma+\gamma'-i\E}
	+\frac{1}{4}\frac{(\gamma/\gamma'+1)^2}{\gamma-\gamma'-i\E}
	-\frac{1}{2}\frac{\gamma^2/\gamma'^2-1}{\gamma-i\E}
,	
	\\
\widehat \pi(\E) &= \frac{\gamma/\gamma'+1}{\gamma-\gamma'-2i(\E-\varepsilon)} - \frac{\gamma/\gamma'-1}{\gamma+\gamma'-2i(\E-\varepsilon)},
\end{align}\end{subequations}and $\widehat{\pi^*}(\E)=[\widehat{\pi}(-\E^*)]^*$
determine the \refb{finite number of poles of the propagator} $\hat{\Pi}(\E)$ listed in Table~\ref{tab:poles}.

It is now straightforward~\cite{Vacchini10, Smirne10} to determine the generator $\G(t)=i\dot\Pi(t) \Pi^{-1}(t)$ and the kernel $\hat\K(\E)=\E \ones-i\hat\Pi^{-1}(\E)$ whose relation has our interest. The spectral decomposition for the generator reads
\begin{align}
\G(t)
	=
	&
	2 i \,  \Re\left( \frac{\dot \pi(t)}{\pi(t)}\right)
	\Big[ \Ket{11}-\Ket{00} \Big] \Bra{11}
	\notag
	\\
	& + i
	\frac{\dot \pi(t)}{\pi(t)}
	\Ket{01}\Bra{01}
	+ i \left(
	\frac{\dot \pi(t)}{\pi(t)}	
	\right)^* \Ket{10}\Bra{10} 
	\label{eq:jc_model_G_t}
	,
\end{align}
whereas for the kernel in the frequency domain it is
\begin{align}
\hat\K(\E)
= &
 \Big( \E-\frac{i}{\widehat{|\pi|^2}(\E)}\Big)
 \Big[ \Ket{11}-\Ket{00} \Big] \Bra{11}
\label{eq:K-frequency}
 \\
&+\Big(\E-\frac{i}{\widehat{\pi}(\E)}\Big) \Ket{01}\Bra{01}
+
\Big(\E-\frac{i}{\widehat{\pi^*}(\E)}\Big)
\Ket{10}\Bra{10}
\notag
.
\end{align}
The eigenvalues of $\hat{\K}$ satisfying $k_{j}(\E_p)=\E_p$ for some $j$
correspond to the poles of $\hat{\Pi}(\E)$ in Table~\ref{tab:poles}.

\subsubsection{Overdamped dynamics ($\gamma \geq 2 \Gamma$)}

\begin{table}[t]
	\centering
	\renewcommand{\arraystretch}{1.1}
	\begin{tabular}{ l  c  l}
		\hline\hline
		Poles $\hat{\Pi}(\E)\phantom{x^{x^{x^{x}}}}$                         &                  & Eigenvalues $\G(\infty)$                           \\
		$\E_0=0$                                        &                  & $g_0=0$                                            \\
		$\E_1= +\varepsilon-i\tfrac{1}{2}(\gamma-\gamma')$ &                  & $g_1= + \varepsilon-i\tfrac{1}{2}(\gamma-\gamma')$ \\
		$\E_2= -\varepsilon-i\tfrac{1}{2}(\gamma-\gamma')$ &                  & $g_2=-\varepsilon-i\tfrac{1}{2}(\gamma-\gamma')$   \\
		$\E_3= -i(\gamma-\gamma')$                         &                  & $g_3=-i(\gamma-\gamma')$                           \\
		$\E_4= +\varepsilon-i\tfrac{1}{2}(\gamma+\gamma')$ & $\longleftarrow$ & \emph{Possibly closer to real axis}               \\
		$\E_5= -\varepsilon-i\tfrac{1}{2}(\gamma+\gamma')$ & $\longleftarrow$ & \emph{than $\E_3$!}               \\
		$\E_6= -i\gamma$                                   &                  &  \\
		$\E_7	=-i(\gamma+\gamma')$                          &                  & \\
		\hline\hline
	\end{tabular}
	\caption{\refb{Jaynes-Cummings model:
		Poles of $\hat{\Pi}(\E)$ and eigenvalues of $\G(\infty)$
		using the abbreviation}
		$\gamma' = \sqrt{\gamma(\gamma -2\Gamma)}$.
	\label{tab:poles}}
\end{table}

Even with all explicit expressions in hand,
it is by no means obvious that this model obeys our sampling result~\eq{eq:sampling}
in the stationary limit $t\to \infty$.
We now first verify this noting that our assumption that $\G(\infty)$ exists holds only
for broad spectral densities such that $\gamma \geq 2 \Gamma$.
In this case the real quantity $\gamma' = \sqrt{\gamma(\gamma-2\Gamma)} \leq \gamma$
represents a suppression/enhancement of the decay rates $-\Im \, \E_p$ relative to the value $\gamma$ in Table~\ref{tab:poles}.
\refb{In this overdamped regime
$\lim_{t\rightarrow\infty} {\dot \pi(t)}/{\pi(t)} = 
-\tfrac{1}{2}(\gamma-\gamma')-i\varepsilon$ converges
and the dynamics is CP-divisible~\footnote{The generator can be written in time-dependent GKSL form
	$-i \G(t) = -i[H,\bullet]+ jJ \bullet J^\dag$,
	with $H=\varepsilon \ket{1}\bra{1}$ and jump operator $J=\ket{0}\bra{1}$.
	For $\gamma < 2\Gamma$ the  jump rate $j=- 2 \Re [\dot{\pi}(t)/\pi(t)]$ can be negative by \Eq{eq:phase_jump}
	which is equivalent to CP-divisibility~\cite{Rivas10,Rivas14}.
	For $\gamma \geq 2 \Gamma$ we have $j>0$ by
	${\dot{\pi}(t)}/{\pi(t)} =
	-i\varepsilon -\tfrac{1}{2} \gamma 
	+\tfrac{1}{2}\gamma'
	\tanh ( \tfrac{1}{2}\gamma' t +\tanh^{-1} \frac{\gamma}{\gamma'} )$.
}.
}

Table~\ref{tab:poles} shows that the resulting four \emph{eigenvalues} of $\G(\infty)$ indeed coincide with four of the eight \emph{poles} of $\hat{\Pi}(\E)$
as predicted by \Eq{eq:pole}.
Interestingly, $\G(\infty)$ does \emph{not always} sample the \enquote{slowest} part of the evolution, i.e., the poles with the smallest decay rates, even in this simple model.
Whereas this happens for sufficiently large broadening
$\gamma > \frac{9}{4} \Gamma$,
just before entering the underdamped regime there is a range $2\Gamma < \gamma < \frac{9}{4}\Gamma$,
where two non-sampled poles $\E_{4,5}$ have smaller decay rates than the sampled pole $\E_3$, see Table~\ref{tab:poles}.
\refb{Thus, $\G(\infty)$ is completely determined by the sampling of $\hat{\K}(\E)$ as dictated by \Eq{eq:G-sampling}.
This does not not illustrate the full complexity of the sampling since the right eigenvectors of $\hat\K(\E)$ are frequency independent and thus trivially provide the right eigenvectors~\eq{eq:eigenvector} of $\G(\infty)$.
}

Numerical implementation of the stationary iteration described in \Sec{sec:stationary-iteration} converges in a few steps to the exact stationary generator, which explicitly reads
\begin{gather}
	-i\G(\infty)  =
	- \frac{2\Gamma}{1+\sqrt{1-2\Gamma/\gamma}}
	 \Big[ \Ket{11}-\Ket{00} \Big] \Bra{11}
	\label{eq:G-infty}
\\
	- \Big(+i \varepsilon + \frac{\Gamma}{1+\sqrt{1-2\Gamma/\gamma}} \Big)\Ket{01}\Bra{01}
	\notag \\
	-\Big(-i \varepsilon + \frac{\Gamma}{1+\sqrt{1-2\Gamma/\gamma}} \Big)\Ket{10}\Bra{10}
	\notag
	.
\end{gather}Importantly, we numerically observe this convergence starting from \emph{random}
initial superoperators $X$.
Although other fixed points of \Eq{eq:fixed-point-stationary-b} can be constructed~\footnote{
		(i)~Select any four poles $\E_{s_1},\dots,\E_{s_4}$ from Table~\ref{tab:poles}
		that have \textit{linearly independent} right eigenvectors.
		These are eigenvectors from \emph{different} superoperators $\hat\K(\E_i')$ and thus need not be linearly independent.
		(ii)~From this basis $\Ket{k_{s_1}},\dots,\Ket{k_{s_4}}$ construct a corresponding dual basis $\Bra{\bar k_{s_1}}, \dots, \Bra{\bar k_{s_4}}$.
		(iii)~Construct a fixed point as
		$\G_{s_1\ldots s_4}=\sum_{i=1}^{4} \E_{s_i} \Ket{k_{s_i}} \Bra{\bar k_{s_i}}$.
},
we always find that $\G(\infty)$ is the only stable one.
Due to this remarkable fact, the iterative solution allows one to infer which of the poles are sampled by $\G(\infty)$.
\refb{As mentioned earlier} this can be used to assist the identification of the sampled poles in analytical calculations, which aim to exploit \Eq{eq:sampling}.

\begin{figure}[t]
	\centering
	\includegraphics[width=\linewidth]{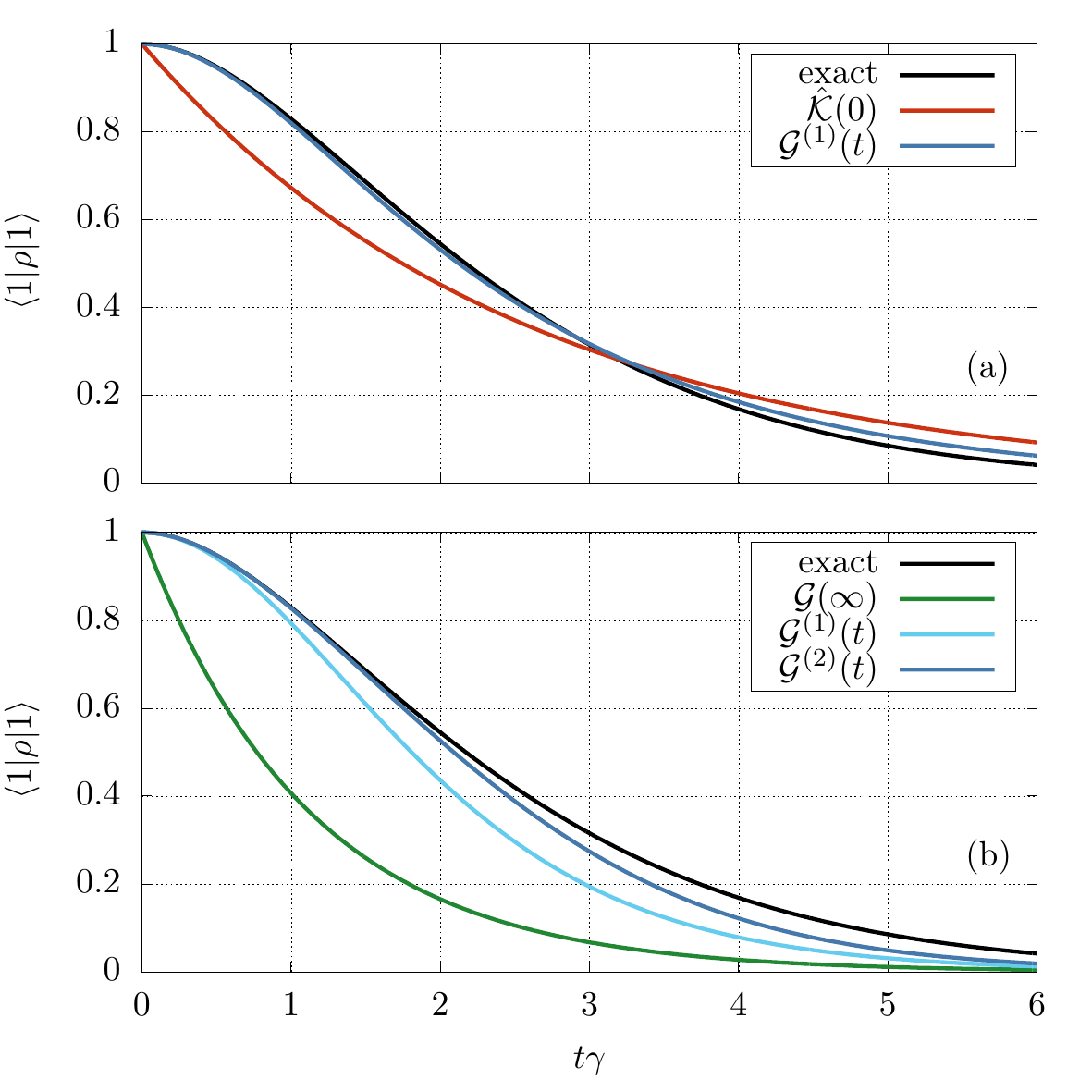}
	\caption{
		\refb{Jaynes-Cummings model,} overdamped regime ${\Gamma}/{\gamma}=0.495$~(${\gamma'}/{\gamma}=0.1$):
		Decay of the probability $\bra{1}\rho(t) \ket{1}$
		for the excited state when it is initially occupied, $\rho(0)=\ket{1}\bra{1}$.
		(a)  Solutions obtained from the Markovian approximation $\dot \rho(t) \approx -i \hat{\mathcal{K}}(0) \rho(t)$ (red) together with the first iteration $\G^{(1)}$ (cyan) of the transient fixed-point equation~\eq{eq:fixed_point_decomposed} starting from $\G^{(0)}=\hat{\mathcal{K}}(0)$. The exact solution is shown in black.
		(b) Solutions obtained from the Markovian approximation $\dot \rho(t) \approx -i \G(\infty) \rho(t)$ (blue) together with two iterations of the transient fixed-point equation, $\G^{(1)}$ (cyan)
		and $\G^{(2)}$ (green), when started from $\G^{(0)}=\G(\infty)$
		\refb{and exact solution (black).}
	}
	\label{fig:Ginf_vs_K0_weak}
\end{figure}

Given the kernel $\hat{\K}(\E)$, one can thus find $\G(\infty)$
by iteration \emph{directly} at stationarity,
avoiding the transient time-dependence of~$\G(t)$.
We plot the resulting semigroup approximation \eq{eq:approx-G} in \Fig{fig:Ginf_vs_K0_weak}(b) and the \emph{different} semigroup~\eq{eq:approx-K}, generated by the exact low-frequency kernel
\begin{align}
	-i \hat{\K}(0) =
	&
	- \frac{\Gamma}{1+\Gamma/(2\gamma)}
	\Big[ \Ket{11}-\Ket{00} \Big] \Bra{11}
	\notag
	\\
	&
	- \left( +i\varepsilon + \frac{\gamma \Gamma}{2(\gamma+i\varepsilon)} \right) \Ket{01}\Bra{01}
	\notag
	\\
	&
	- \left( - i\varepsilon +\frac{\gamma \Gamma}{2(\gamma-i\varepsilon)} \right) \Ket{10}\Bra{10}
	,
	\label{eq:K0}
\end{align}
in \Fig{fig:Ginf_vs_K0_weak} (a).
The $\hat{\K}(0)$ semigroup crosses the exact solution already at intermediate times to approach it from above,
whereas the $\G(\infty)$ semigroup approaches it from below.
Indeed, in the overdamped regime the occupation decay rate of \Eq{eq:G-infty} is always larger than that of \Eq{eq:K0}.
As expected, both semigroups have problems with the initial nonlinear time-dependence on the scale $\gamma^{-1}$
set by the reservoir bandwidth~\eq{eq:jc_model_lorentzian}.
Only in the wide-band limit $\gamma \to \infty$
the exact evolution is a semigroup, which in this case is generated by $\G(\infty)=\hat{\K}(0)$.

We have also implemented the functional iteration $\G^{(n)}(t)$ for the transient generator explained in \Sec{sec:time_dependent_iteration}, using \Eq{eq:fixed_point_decomposed} with $\K_\l = -i[H,\bullet]$ and $H=\varepsilon d^\dag d$.
In \Fig{fig:Ginf_vs_K0_weak} (a) and (b) we additionally show the evolutions generated by the approximate $\G^{(n)}(t)$ starting from the initial function $\G^{(0)}(t)=\hat\K(0)$ and $\G(\infty)$, respectively.
Like the semigroups, each approximation approaches the exact stationary state at large times.
However, contrary to the semigroups, each iteration is also very accurate at short times, see~\Eq{eq:explicit_first_iteration}.
These two constraints enforce rapid convergence at intermediate times
throughout the overdamped parameter regime: in \Fig{fig:Ginf_vs_K0_weak}~(a) and (b)
we did not plot the $n=2$ and $n=3$ approximations, respectively, since they are hard to distinguish from the exact solution. Thus, \Fig{fig:Ginf_vs_K0_weak}~(a) shows that \Eq{eq:here}, based solely on \refb{one iteration of} the \emph{time-nonlocal} memory kernel, already provides a remarkably accurate representation of the \emph{time-local} generator.

\subsubsection{Underdamped dynamics ($\gamma < 2\Gamma$)}

\refb{For  narrow spectral density, $\gamma < 2\Gamma$,
the evolution becomes underdamped and non-divisible. The function
}\begin{align}
	\pi(t) = e^{-i\varepsilon t}
	e^{-\gamma t/2} \Big[ \cos \Big(\frac{\Omega t}{2} \Big) + \frac{\gamma}{\Omega} \sin\Big(\frac{\Omega t}{2}\Big) \Big]
	\label{eq:alpha-underdamped}
\end{align}
now oscillates with frequency $\Omega \equiv -i\gamma'=\sqrt{\gamma(2\Gamma-\gamma)}$
with roots located at
$
t_n = \tfrac{2\pi}{\Omega} \big( n - \tfrac{1}{\pi} \arctan\frac{\Omega}{\gamma} \big)
$.
This qualitative change of $\pi(t)$ has two consequences.

First, the time-local generator $\G(t)$ \emph{by itself} exhibits singularities as function of time for every $t=t_n$ [\Eq{eq:jc_model_G_t}].
\refb{These dynamics with singular generators have recently received renewed attention~\cite{Chruscinski18,Chakraborty19,Chakraborty20}, even though they have been noted long ago~\cite{BreuerPetruccione}.
Importantly these singularities}
are not spurious,
noting that the product $\G(t)\Pi(t)$ remains finite even at $t=t_n$.
\refb{In fact they are
}physically meaningful:
by identifying the divergent matrix elements of $\G(t)$ one can already infer at which times the solution \refb{of the Jaynes-Cummings model} will be an entanglement breaking map~\cite{Yu04,Almeida07,Laurat07,Yu09}, $\Pi(t_n)=\Ket{00}\Bra{\one}=\ket{0}\bra{0} \Tr \bullet$.

\begin{figure}[t]
	\centering
	\includegraphics[width=\linewidth]{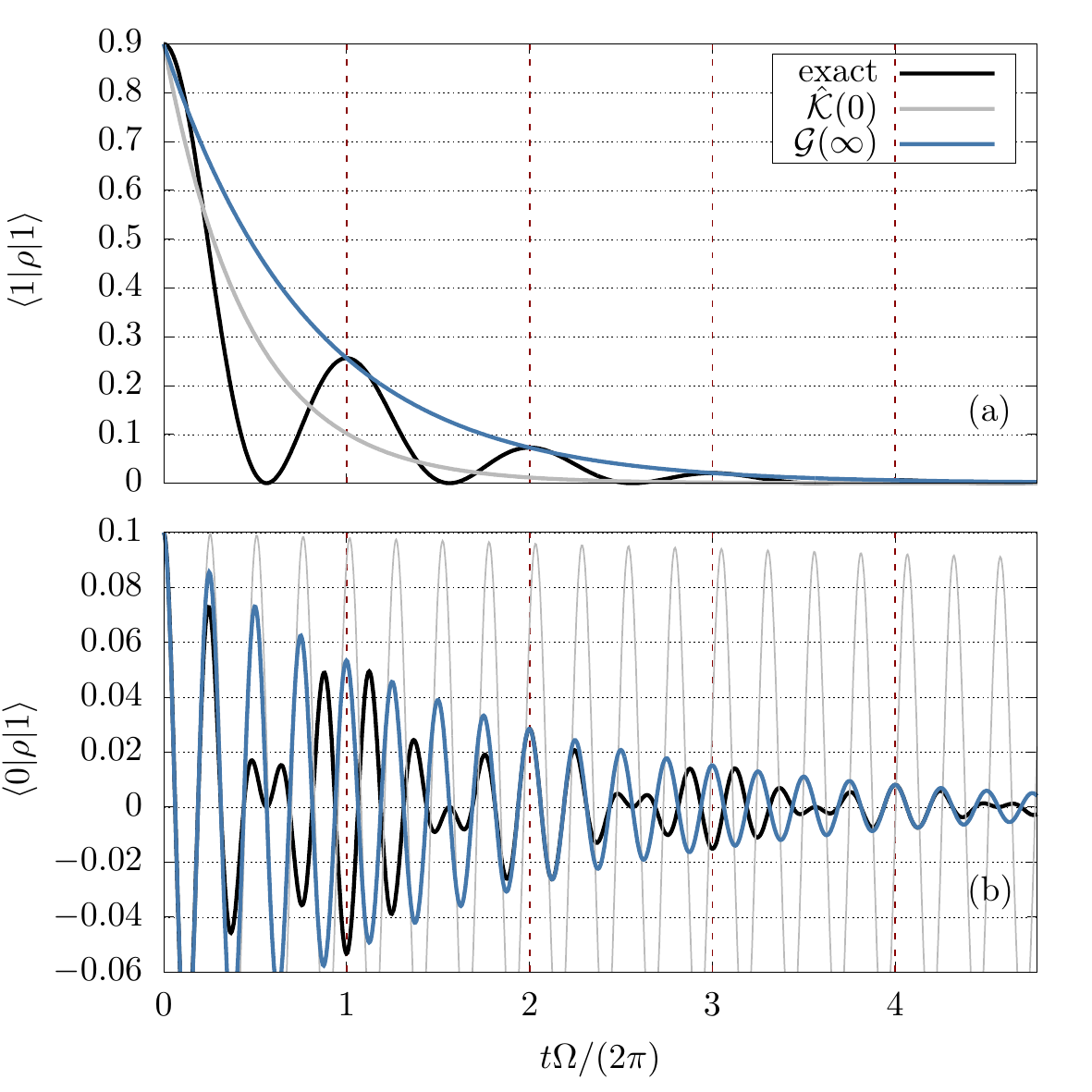}
	\caption{
		Jaynes-Cummings model, underdamped regime ${\Gamma}/{\gamma}=13$ ($\Omega/\gamma=5$) and $\varepsilon=20$:
		(a)~Decay of the excited state occupation $\bra{1}\rho(t) \ket{1}$ and
		(b)~decay of the real part of the coherence $\bra{0}\rho(t)\ket{1}$.
		The initial state is $\rho(0)=0.1 \big[ \Ket{00} + \Ket{01} + \Ket{10}\big]+0.9 \Ket{11}$.
		Shown are the solution for the Markovian approximations $\dot \rho(t) \approx -i \G(\infty) \rho(t)$ [blue]
		with generator obtained by iteration of the stationary fixed point equation~\eq{eq:fixed-point-stationary}, and $\dot \rho(t) \approx -i \hat{\K}(0) \rho(t)$ [gray].
		The exact solution is shown in black.
		\label{fig:Ginf_vs_K0_strong}
	}
\end{figure}

A second consequence is that	
the stationary limit of $\G(t)$ \emph{by itself} does not exist, even though
the stationary propagator does converge, $\lim_{t\to \infty}\Pi(t)=\Ket{00}\Bra{\one}$,
and  the low-frequency memory kernel $\hat{\K}(0)$ is well defined.
\refb{
Irrespective of how generic both these complications are,
they present perhaps the most crucial challenge to \emph{any} time-local approach. It is well known, for example, that perturbative calculations of $\G(t)$ cannot venture beyond the first singularity on the time axis~\cite{Breuer99,BreuerPetruccione}.
In this sense the model presents a worst-case test for both variants of the fixed point iteration.

\refb{The stationary iteration [\Sec{sec:stationary-iteration}] is simply expected to fail since it relies on the convergence of $\G(t)$ for $t\to \infty$}.
Nevertheless,
it is interesting to explore what happens.
Indeed, }the stationary iteration for $\G^{(n)}(\infty)$ does not converge anymore with $n$.
\refb{However,
}$\G$ is always block diagonal
and we observe that the iterations for the generator on the occupation subspace $\Ket{00},\Ket{11}$ converge to
\begin{align}
\lim_{n\to \infty} \G^{(n)}_{o}(\infty)
= -i \gamma  \big[ \Ket{11}-\Ket{00} \big] \Bra{11}
,
\label{eq:jc_model_g_iterated}
\end{align}
whereas the generator $\G^{(n)}_{c}$ on the subspace $\Ket{01},\Ket{10}$ of the coherences oscillates indefinitely with $n$.
In \Fig{fig:Ginf_vs_K0_strong}a we plot the time-evolution of occupations obtained from the semigroup approximation constructed from \Eq{eq:jc_model_g_iterated}.
In contrast to the semigroup generated by the well-defined $\hat{\K}(0)$,
it gives an accurate envelope the decay of the excited state,	
even in the strongly underdamped limit,
$\gamma \ll \Gamma$ where $\Omega \approx \sqrt{2\Gamma\gamma} \gg \gamma$.

\begin{figure}[t]
	\centering
	\includegraphics[width=\linewidth]{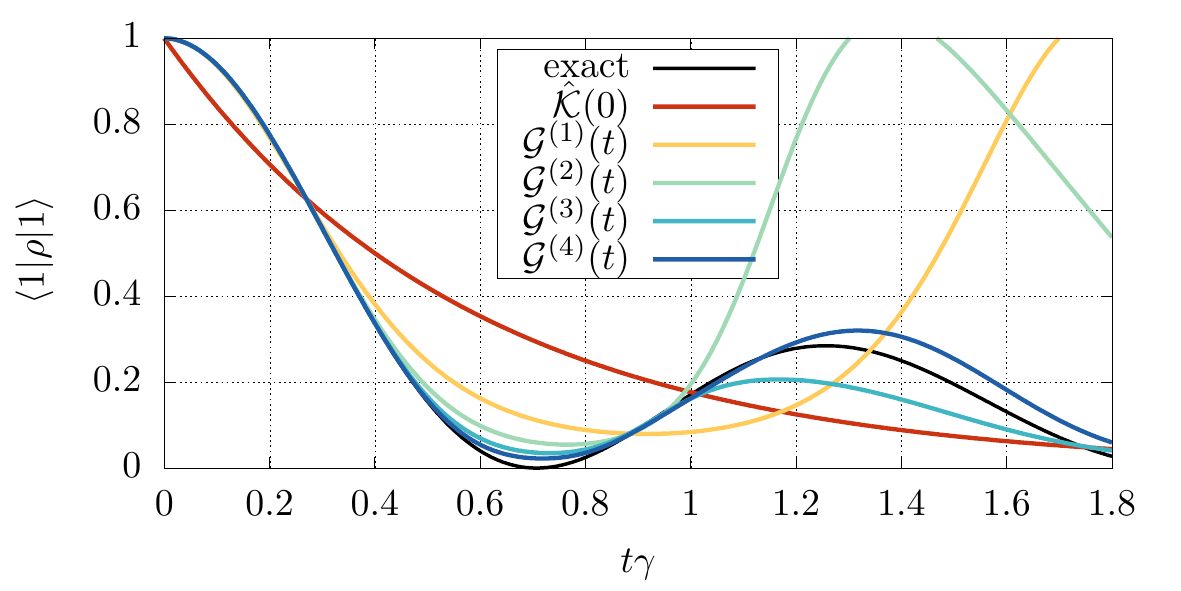}
	\caption{
		\refb{
			Jaynes-Cummings model, underdamped regime:
			Decay of the probability $\bra{1}\rho(t) \ket{1}$ obtained
			from generators of the transient iteration for
			${\Gamma}/{\gamma}=13$ and $\varepsilon=20$ using the starting point $\hat\K(0)$.
			\label{fig:underdamped_transient_iteration}
		}
	}
\end{figure}

The converged part of the iteration 
can in fact be related to a regularization of $\lim_{t\to \infty} \G(t)$.
Noting that
\begin{align}
\frac{\dot{\pi}(t)}{\pi(t)} = -i\varepsilon -\tfrac{1}{2} \gamma
 - \tfrac{1}{2}\Omega \tan\Big( \tfrac{1}{2}\Omega t
 - \arctan \frac{\gamma}{\Omega} \Big)
\label{eq:phase_jump}
\end{align}
we see that a principal-value time-average over one period
amounts to replacing
$\frac{\dot{\pi}(t)}{\pi(t)} \to
-i\varepsilon -\tfrac{1}{2} \gamma$.
This gives a regularized stationary limit for the generator,
\begin{align}
\G(\infty)_\text{reg} =&
-i\gamma \big[ \Ket{11}-\Ket{00} \big] \Bra{11}
\label{eq:G-infty-reg} \\
&-(-\varepsilon + \tfrac{1}{2} i \gamma)  \Ket{01}\Bra{01}
-(\varepsilon + \tfrac{1}{2} i \gamma)  \Ket{10}\Bra{10}
.
\notag
\end{align}
which coincides with the numerically converged block \eqref{eq:jc_model_g_iterated} of the iteration.
\refb{The value of the coherence block exposes a key complication of the exact evolution of this model.}
In \Fig{fig:Ginf_vs_K0_strong}b, we show that the semigroup constructed from $\G(\infty)_\text{reg}$
describes the decay \refb{\emph{and} oscillation} of the coherences accurately in the center of every \emph{even} time interval. However, it is \emph{also} accurate  up to the \emph{sign} in every \emph{odd} interval. The intermediate $\pi$-phase jumps occurring in the exact solution are caused by the divergences of the generator at times $t_n$.
\refb{The stationary
}fixed-point iteration may thus still be useful beyond the limitations we \refb{assumed} in the present paper.

\refb{
Finally, we consider how the transient fixed-point iteration [\Sec{sec:time_dependent_iteration}] deals with the time-singularities in this model. In \Fig{fig:underdamped_transient_iteration}
we show how the occupations, starting from the semigroup approximation generated by $\hat{\K}(0)$, converge to the exact solution. The first two iterations only improve the solution before the first singularity and even become unphysical at larger times. However, the following iterations also converge beyond the first singularity. The fifth iteration (not shown) is indistinguishable from the exact solution in the shown time interval. More iterations are required to converge the solution in a larger time interval also including the second singularity.  The success of our iteration strategy starting from the memory kernel $\K$ highlights its difference to perturbation theory, which always fails in capturing dynamics beyond a singularity~\cite{BreuerPetruccione}. Thus even time-singular generators can be locally stable fixed points of the functional $\hat\K$. Finally we note that the fixed point iteration for the coherences is more challenging, but we do not exclude that this is only a numerical challenge.

}

 \refb{\begin{table}[t]
	\centering
	\renewcommand{\arraystretch}{1.1}
	\begin{tabular}{ l  c  l}
		\hline\hline
		Poles $\hat{\Pi}(\E)\phantom{x^{x^{x^{x}}}}$   &              & Eigenvalues $\G(\infty)$                       \\
		$\E_0=0$                                       &              & $g_0=0$                                        \\
		$\E_1= +(\varepsilon-\mu)-i\tfrac{1}{2}\Gamma$ &              & $g_1= + (\varepsilon-\mu)-i\tfrac{1}{2}\Gamma$ \\
		$\E_2= -(\varepsilon-\mu)-i\tfrac{1}{2}\Gamma$ &              & $g_2= -(\varepsilon-\mu)-i\tfrac{1}{2}\Gamma$  \\
		$\E_3= -\Gamma$                                &              & $g_3=-\Gamma$                                  \\
		$\E_{4+2n}= \E_1 -i \pi T(2n+1)$           & $\leftarrow$ & \emph{Possibly closer to real axis}            \\
		$\E_{5+2n}= \E_2 -i \pi T(2n+1)$           & $\leftarrow$ & \emph{than $\E_3$!}
		\\
		\hline\hline
	\end{tabular}
	\caption{Resonant level model: $n=0,1,2,\ldots$
		\label{tab:poles-rlm}}
\end{table}

\subsection{Finite-temperature resonant level model\label{sec:rlm}}

We complement the above by an analysis of the fermionic resonant level model.
Although its generator has no time-singularities,
its time-dependent algebraic structure provides a challenge complementary to the previous model.
The Hamiltonian is formally identical to that of the Jaynes-Cummings model
except that the reservoir operators are fermionic,
$\{ b_\omega,b_{\omega'}^\dag \}= \delta(\omega-\omega') \one$.
Also, we consider the reservoir at temperature $T$ and chemical potential $\mu$ coupled with an energy independent spectral density $\Gamma=\text{const.}$
This is the most basic model of transient electron tunneling from a localized state.
Even though it ignores interaction effects, its propagator is feature rich.
This was noted in recent work~\cite{Reimer19b}, but the nontrivial relations between $\K$ and $\G$ and their spectra noted below were overlooked. The diagonal representation of $\Pi$ reads
(\Ref{Reimer19b}, Eq. (E1))
\begin{table*}
	\caption{\label{tab:vectors}
		Sampling of memory kernel $\hat{\K}(\E)$ by the stationary generator $\G(\infty)$ [\Eq{eq:sampling}].
		Left columns: for each \emph{different} superoperator $\hat{\K}(g_i)$
		we list \emph{one} pole-eigenvalue with its left and right eigenvector.
		Right columns: collecting the right eigenvectors from $\hat{\K}(g_i)$ and biorthonormalizing we construct  the left eigenvectors $\Bra{g'_i}$. Row $i=1,2$ corresponds to $\eta = \pm$.
	}.
	\renewcommand{\arraystretch}{1.5}
	\begin{ruledtabular}
		\begin{tabular}{crcl rcl}
			& \emph{(iv) Don't copy these!} & $\hat{\K}(g_i)$ & \emph{(i) Copy these} & 
			\emph{(iii) Biorthogonalize}
			& $\G(\infty)$ & \emph{ (ii) Collect here}
\\
			$i$
			& $\Bra{\hat{k}'_{j_i}(g_i)}$
			& $\hat{k}_{j_i}(g_i)$
			&$\Ket{\hat{k}_{j_i}(g_i)}$
			&$\Bra{g_{i}'}$
			&$g_i$
			&$\Ket{g_i}$
			\\
			\hline
			$0$ & $\Bra{\one}$ &
			$0$ &
			$\frac{1}{2} \big[ \Ket{\one}+\hat{\k}(i\tfrac{\Gamma}{2}) \Ket{(-\one)^N} \big]$ &
$\Bra{\one}$ &
			$0$ &
			$\frac{1}{2} \big[ \Ket{\one}+\hat{\k}(i\tfrac{\Gamma}{2}) \Ket{(-\one)^N} \big]$ 
			\\
$1,2$ &
			$\downarrow$\,\,\quad $\Bra{d_\eta^\dag}$ &
			$- \eta \varepsilon -i\frac{1}{2}\Gamma$ &
			$\Ket{d_\eta^\dag}$ &
			$\downarrow$\,\,\quad 
			$\Bra{d_\eta^\dag}$ &
			$ - \eta \varepsilon -i\frac{1}{2}\Gamma $ &
			$\Ket{d_\eta^\dag}$ 
			\\
			$3$ & $\frac{1}{2} \big[ \Bra{(-\one)^N} - \hat{\k} \big(- i\tfrac{\Gamma}{2} \big) \Bra{\one} \big]$ &
			$-i\Gamma $ &
			$\Ket{(-\one)^N}$ &
			$\frac{1}{2} \big[ \Bra{(-\one)^N}-\hat{\k}(i\tfrac{\Gamma}{2}) \Bra{\one} \big]$ &
$-i\Gamma$ &
			$\Ket{(-\one)^N}$ 
\end{tabular}
	\end{ruledtabular}
\end{table*}
\begin{subequations}
	\begin{align}
	&
	\Pi(t) =
	\sum_{\eta=\pm} e^{(i\eta\varepsilon-\tfrac{1}{2}\Gamma)t}
	\Ket{d^\dag_\eta} \Bra{d^\dag_\eta}
	\\
	&
	+ \tfrac{1}{2} \left[ \Ket{\one} +
	\p(t)\Ket{(-\one)^N} \right]
	\Bra{\one}
	\\
	&
	+ e^{-\Gamma t} \tfrac{1}{2} \Ket{(-\one)^N} \left[ \Bra{(-\one)^N} -
	\p(t) \Bra{\one} \right],
	\end{align}\label{eq:pi-rlm}\end{subequations}where $d_+ \equiv d^\dagger$, $d_- \equiv d$, $\Ket{O}\equiv O$ and $\Bra{O} \equiv \text{tr} (O^\dag\bullet)$ for an operator $O$.
In contrast to the Jaynes-Cummings model its eigenvectors depend on time through the function
\begin{align}
\p(t)=
\sum_{\eta=\pm} \eta \, \Im
& \Big[
\frac{
	e^{-(\pi T+i\epsilon)t}
}
{ \pi \sinh(\Gamma t/2) }
 \Phi(e^{-2\pi T t},1,   \tfrac{1}{2} + \tfrac{i \epsilon+ \eta \Gamma/2 }{2\pi T}) 
\notag
\\
&
+
\frac{
	e^{\eta \Gamma t/2} 
}
{ \pi \sinh(\Gamma t/2) }
\Psi(\tfrac{1}{2} + \tfrac{i \epsilon+ \eta \Gamma/2 }{2\pi T})
\Big]
\end{align}
involving Lerch ($\Phi$) and digamma ($\Psi$) functions with $\epsilon = \varepsilon-\mu$.
This richer structure is also reflected by the analytic properties of the propagator
(\Ref{Reimer19b}, App. D)
\begin{subequations}
	\begin{align}
		&
		\hat{\Pi}(\E) =
		\sum_{\eta=\pm} \frac{i}{\E+\eta\varepsilon+i\frac{\Gamma}{2}}
		\Ket{d^\dag_\eta} \Bra{d^\dag_\eta}
		\\
		&
		+ \frac{i}{\E} \tfrac{1}{2} \left[ \Ket{\one} +
		\hat{\k}\left(\E+i\tfrac{\Gamma}{2}\right)\Ket{(-\one)^N} \right]
		\Bra{\one}
		\\
		&
		+ \frac i{\E+i\Gamma} \tfrac{1}{2} \Ket{(-\one)^N} \left[ \Bra{(-\one)^N} -
		\hat{\k} \left(\E +i\tfrac{\Gamma}{2}\right) \Bra{\one} \right]
	\end{align}\label{eq:pi-frequency-level}\end{subequations}expressed in the Laplace transform $\hat{\k}(\omega) \equiv \int_0^\infty dt e^{i \omega t}\k(t)$ of
$\k(t) \equiv 2T {\sin[(\varepsilon-\mu)t]}/{\sinh[\pi T t]}$.
Its poles, listed in Table~\ref{tab:poles-rlm}, include two infinite series for $T>0$, which merge into branch cuts as $T \rightarrow 0$.

The generator $\G(t)=i\dot\Pi(t) \Pi^{-1}(t)$ (\Ref{Reimer19b}, Eq. (B14))
\begin{align}
&
\G(t) =
\sum_{\eta=\pm} \big( -\eta\varepsilon-i\tfrac{1}{2} \Gamma \big)
\Ket{d^\dag_\eta} \Bra{d^\dag_\eta}
\notag\\
&
-i \Gamma \tfrac{1}{2} \Ket{(-\one)^N}
\Big[ \Bra{(-\one)^N} - \g(t) \Bra{\one} \Big]
\label{eq:G-rlm}\end{align}
is obtained with $\g(t)=\int_0^t ds e^{-\frac{1}{2}\Gamma s} \k(s)$, which is related to $\p(t)=\Gamma/(1-e^{-\Gamma t}) \int_0^t ds e^{-\Gamma(t-s)} \g(s)$.
The evolution changes its Markovian character from CP divisible ($|\g(t)|\leq 1$) close to resonance
to non-divisible sufficiently
far from resonance.
The kernel $\hat\K(\E)=\E \ones-i\hat\Pi^{-1}(\E)$ can be expressed as (\Ref{Reimer19b}, (D13))
\begin{align}&
\hat{\K}(\E) =
\sum_{\eta=\pm} \big( -\eta\varepsilon-i\tfrac{1}{2} \Gamma \big)
\Ket{d^\dag_\eta} \Bra{d^\dag_\eta}
\notag\\
&
-i \Gamma \tfrac{1}{2} \Ket{(-\one)^N}
\Big[ \Bra{(-\one)^N} - \hat{\k} \big(\E +i\tfrac{1}{2}\Gamma \big) \Bra{\one} \Big]
.
\label{eq:K-frequency-level}\end{align}Unlike the Jaynes-Cummings model, none of these superoperators commute with themselves at different time/frequency/parameter values (on which their eigen\emph{vectors} depend) nor with each other (since $\p(t), \k(t),\g(t)$ all differ).
We now show that, nevertheless, the sampling relation~\eqref{eq:sampling} explicitly holds.

Table~\ref{tab:poles-rlm} shows that the four eigenvalues of $\G(\infty)$ indeed coincide with four of the poles of $\hat \Pi(\E)$, which coincide with the four frequency-independent eigenvalues of $\hat\K(\E)$.
However, the propagator $\hat{\Pi}(\E)$ has infinitely many more poles $\{ \omega_{n} \}_{n \geq 4}$
which arise from the function
$\hat{\k} \big(\E +i\tfrac{1}{2}\Gamma \big)$
located in the eigenvectors of $\hat\K(\E)$.
These are not sampled as explained after \Eq{eq:eigenvector}.
For $T \leq \Gamma/(2\pi)$ some of these non-sampled poles lie \emph{in between} the sampled poles $\E_{1},\E_{2}$ and $\E_3$ and form branch cuts as $T \to 0$.

In Table~\ref{tab:vectors} we illustrate how $\G(\infty)$ also nontrivially samples the eigen\emph{vectors} of $\hat{\K}(\omega)$ as follows:
(i) We collect one right eigenvector from each of the four \emph{different} superoperators
$\hat{\K}(0)$, $\hat{\K}(\pm \varepsilon -i\tfrac{1}{2}\Gamma)$ and $\hat{\K}(-i\Gamma)$
.
(ii) This gives four right vectors
$\Ket{\hat{k}_{j_i}(g_i)}=\Ket{g_i}$.
(iii) From this set one \emph{algebraically} constructs a set of biorthonormal covectors $\Bra{g_i'}$.
This way we remarkably obtain the left and right eigenvectors of $\G(\infty)$ as given by
\Eq{eq:G-rlm} using the \emph{analytic} property $\g(\infty)=\hat{\k}(i\tfrac{1}{2}\Gamma)$.
Note in particular that one would \emph{not} obtain the correct \emph{left} eigenvectors of $\G(\infty)$ by naively sampling the left pole-eigenvectors of the kernels.
For eigenvalue $g_3=-i\Gamma$ a difference arises as indicated by the two arrows in Table~\ref{tab:vectors}.

We observe that for the resonant level model each eigenvalue-pole is sampled precisely once by $\G(\infty)$.
Combined with
the mere assumption that $\G(\infty)$ is diagonalizable the sampling relation~\eq{eq:sampling} thus completely determines this superoperator, because it exhausts the number of eigenvalue-poles ($d^2=4$).

For the resonant level model the numerical stationary iteration [\Sec{sec:stationary-iteration}] starting from any initial $\G^{(0)}(\infty)$ also converges to the exact stationary generator.
This holds for \emph{all} parameters of the model.
Strikingly, using $\G^{(0)}(\infty)=\hat{\K}(0)$ as a starting point the iteration terminates right away
at the \emph{zeroth} iteration, implying an exact relation~\footnote{This relation should not be misunderstood as saying that \enquote{$\G(\infty)$ only samples $\K(\omega)$ at $\omega=0$}and that there is something trivial about the sampling: keeping only $\omega=0$ in \Eq{eq:sampling} would give $\G(\infty) =0$.
	Instead, this relation is a nontrivial statement about the memory kernel of this model: sampling $\hat{\K}(\E)$ according to \Eq{eq:sampling} at eigenvalue-poles --including ones at \emph{nonzero} frequencies and \emph{ignoring} eigenvector poles\,/\,branch cuts -- exactly reproduces the \emph{zero}-frequency kernel~$\hat{\K}(0)$.
}\begin{align}
	\G(\infty) = \hat{\K}(0).
	\label{eq:trivial}
\end{align}
One verifies the relation~\eq{eq:trivial} by comparing \Eqs{eq:K-frequency-level} and \eq{eq:G-rlm} again using $\g(\infty)=\hat{\k}(i\tfrac{1}{2}\Gamma)$.

\begin{figure}[t]
	\centering
	\includegraphics[width=\linewidth]{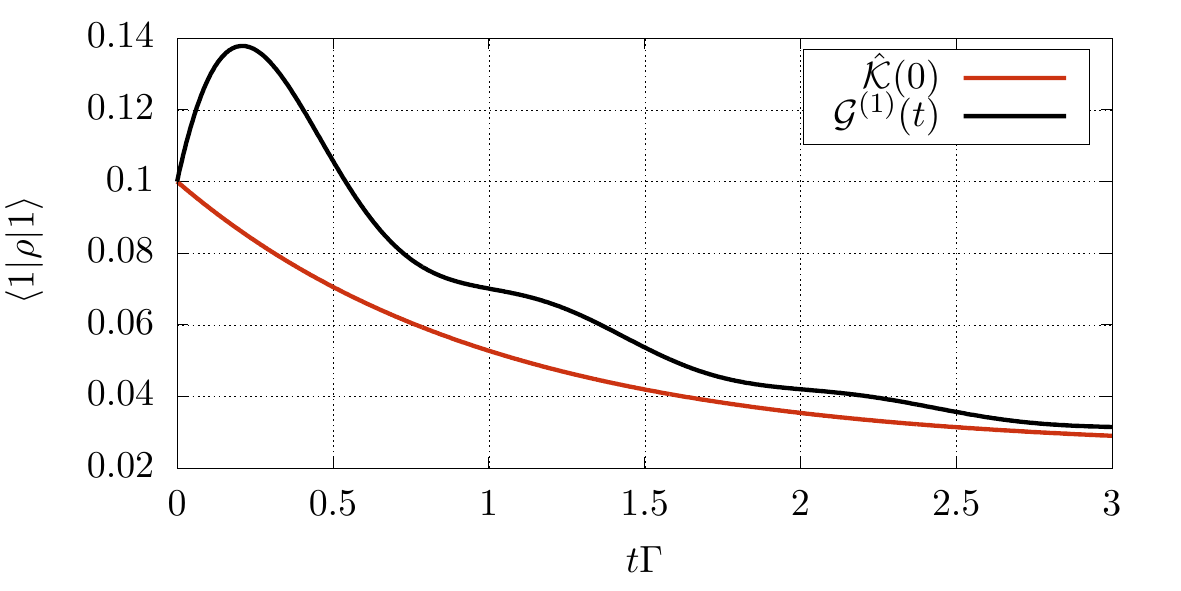}
	\caption{
		\refb{
			Decay of the occupation $\bra{1}\rho(t) \ket{1}$ in the RLM for $\varepsilon-\mu=2\pi\Gamma$, $T=0.1 \cdot \Gamma/(2\pi)$
			obtained from $\G^{(0)}(t)=\hat{\K}(0)$ and $\G^{(1)}(t)=\G(t)$.
			The former corresponds to a Markov semigroup approximation using $\hat{\K}(0)$, which is never able to describe the initial growth of the occupation \emph{away} from the stationary value. When used as initial guess in the transient iteration the exact solution is recovered by the fixed-point equation~\eq{eq:fixed-point} after a single step.
			\label{fig:rlm_iteration}
		}
	}
\end{figure}

The transient iteration [\Sec{sec:time_dependent_iteration}]
starting from the constant ansatz $\G^{(0)}(t) = \hat{\K}(0)$ does not terminate immediately,
because the evolution is not a semigroup.
However, for this ansatz the \emph{first} transient iteration does give the exact solution,
$\G^{(1)}(t)= \int_{0}^{t} ds \K(t-s)e^{i\hat{\K}(0)(t-s)} = \G(t)$, again for all parameters of the model.
This reflects an exact relation [\Ref{Reimer19b}, Eqs. (52a), (D15)]
\begin{align}
\G(t) = \int_{0}^{t} ds \K(t-s)
\label{eq:G-simple}
,
\end{align}
which for $t \to \infty$ again implies~\Eq{eq:trivial}.
In \Fig{fig:rlm_iteration} we show the time-dependence of the occupations for the zeroth and first iteration.
Unlike the Jaynes-Cummings model, the level initially actually fills up more before decaying to the empty stationary state,
an effect caused by time-dependence of eigenvectors of $\Pi(t)$ [\Eq{eq:pi-rlm}].
This reentrant behavior is completely produced in one step by $\G^{(1)}(t)$ from the Markov semigroup approximation $\G^{(0)}=\hat{\K}(0)=\G(\infty)$,
which can never capture an initial growth in the \enquote{wrong direction}.

In fact, any trace-preserving constant ansatz $\G^{(0)}(t) = X$ gives the exact solution after one iteration, $\G^{(1)}(t) = \G(t)$
as shown in \App{app:iteration_terminates}.
Furthermore,
starting from an arbitrary time-constant superoperator, $\G^{(0)}=X \neq 0$,
the \emph{second} transient iteration always reaches the fixed point, because the first iteration $\G^{(1)}(t)$ produces a trace-preserving generator [\Eq{eq:iteration-TP}].
That two iterations suffice for \emph{all} parameters of the model is remarkable since this includes the extended parameter regime where the level is sufficiently off-resonant and the evolution is not CP-divisible~\cite{Reimer19b,Bruch20a}.
For comparison, in the  underdamped regime where the Jaynes-Cummings model is non-CP-divisible many iterations are required [\Fig{fig:Ginf_vs_K0_strong}].
The termination of the fixed-point iteration
is closely related to truncations of (renormalized) coupling expansions for $\K(t)$ and $\Pi(t)$ for fermionic models,
which occur in the absence of \emph{interactions} and for energy independent coupling~\cite{Saptsov12,Saptsov14,Reimer19b}.
This indicates that the number of iterations is related to dynamically generated many-body effects.

 }

 \refall{
\section{Summary and outlook\label{sec:summary}}
}
We have found the general connection
between two canonical approaches to the dynamics of open quantum systems,
the time-local and time-nonlocal quantum master equation.
This relation extends the response function of an open system -- the frequency-domain memory kernel $\hat \K(\E)$ --
to a functional mapping of \emph{superoperator-functions of time}
of which the generator
is a \emph{fixed point}:
$\G(t,t_0)=\hat{\K}[\G](t,t_0)$.
The fixed-point property expresses
that the generator is a characteristic \enquote{frequency} of the evolution produced by the memory kernel. This is very similar to how pole frequencies characterize the response of linear systems in physical sciences and engineering~\cite{Oppenheim83}.
In our general quantum setting, we showed how the fixed-point equation provides a self-consistent solution of the complicated time-domain gradient expansion.
Interestingly, this also revealed a connection of the time-convolutionless approach to a Moyal formulation of quantum theory of \emph{open} systems.

We obtained several general insights into the role of the frequency dependence of the memory  kernel.
We precisely determined how  the stationary generator $\G(\infty)$ samples the \emph{right} eigenvectors and eigenvalues of the memory kernel $\hat{\K}(\E)$ at zero \emph{and nonzero} characteristic frequencies of the evolution.
The sampled  frequencies form a finite subset of the exact poles of the frequency-domain evolution as obtained by the Laplace-resolvent method in the time-nonlocal approach.
Remarkably, knowing only the location of these poles in the complex plane in principle suffices to \emph{completely construct} the stationary generator $\G(\infty)$ from the memory kernel, significantly simplifying analytical calculations.
This generator may also be obtained numerically by iterating the stationary fixed-point equation~\eq{eq:fixed-point-stationary}.

Similarly, the full transient generator may be obtained from the memory kernel by iterating the functional fixed-point equation~\eq{eq:fixed-point}.
At each iteration the approximate generator is both initially \emph{and} asymptotically accurate.
\refall{
Importantly, this iteration strategy only works if the generator is a
locally \emph{stable} fixed point of the kernel functional
and we have shown that even time-singular generators can be locally stable.
We also showcased an evolution with time-dependent eigenvectors
whose generator is exactly found after at most two iterations.
Since our results apply quite generally and can be tailored to both
numerical~\cite{Cohen11,Kidon18} and
analytical~\cite{Schoeller09,Pletyukhov12a,Schoeller18} applications,
they seem relevant to the challenging problems of strongly interacting
open quantum systems dominated by nonperturbative dissipation and memory
effects.
Altogether, this provides new starting points for combining
well-developed memory kernel formalisms to access the advantages of a
time-local description.
We conclude by outlining several such applications.

\emph{(Non-)Markovianity and microscopic models.
}As mentioned in the introduction,
the most obvious application lies in the study of
the divisibility properties of the dynamics, which are directly
accessible via the time-local generator $\G$.
From its time-dependent canonical form~\cite{Hall14} one can extract
both the jump rates (characterizing CP-divisibility)
and the jump operators (characterizing
P-divisibility)~\cite{Wissmann15,Bae2016},
encompassing different degrees of (non-)Markovianity~\cite{Chruscinski15}.
However, the impressive progress in understanding of dynamics and information in
general terms contrasts with the enduring limitation
of their application to \emph{microscopic} models.
Our key result \eq{eq:fixed-point} provides a new path from accurate
memory kernels calculated within the well-developed time-nonlocal
formalism to the time-local description required for these
problems.

\emph{Geometry, topology and transport in open systems.
}Similarly, the study of geometric~\cite{Sarandy05,Sarandy06} and
topological phases~\cite{Li14,Riwar19} hinges on a time-local description.
Starting from the time-local QME,
already in the leading adiabatic approximation to the dynamics
\begin{gather}
\Ket{ \rho(t) } =
\mathcal{T}_{\leftarrow} e^{-i \int_0^t \G(t)} \Ket{\rho(0)}
\label{eq:adiabatic_approx}
\\
\approx
\sum_i \Ket{g_i(t)}
\,
\Braket{\bar{g}_i(t)}{\rho(0)}
\, e^{\int_{0}^t ds [\,  -i g_i(s) -
	\Braket{\bar{g}_i(s)}{\partial_s g_i(s)} \, ] }
\notag
\end{gather}
one finds that the \emph{time-instantaneous} eigenvalues and
eigenvectors  of $\G(t)$ --\emph{not} those of the memory kernel
$\K(t)$-- determine the dynamic and geometric phase of the open system,
respectively.
For stationary driving one can expand these quantities around their
stationary values with \emph{parametric} time dependence.
Our stationary fixed point equation~\eq{eq:fixed-point-stationary}
describes how these required quantities are nontrivially related to
eigenvalues and eigenvectors of the more accessible memory kernel $\K$.
As illustrated in \Sec{sec:summing}, this procedure
automatically includes the \emph{full} memory-expansion of the
kernel~\cite{Splettstoesser06}.

These geometric phases of \emph{state} evolution
subsume~\cite{Pluecker17b} those of
observables~\cite{Ning92,Landsberg92,Landsberg93,Pluecker17} and their
full-counting statistics~\cite{Sinitsyn07PRB,Sinitsyn09} as a
special case by incorporating an ideal counter into the state
evolution~\cite{Nazarov03b,Schaller09}.
This immediately implies that our fundamental
relation~\eq{eq:fixed-point} for state evolution translates to a
nontrivial relation between the time-nonlocal kernel and time-local
generator that govern the dynamics of the \emph{full counting
	statistics}, i.e., the \emph{transport} equations.
The FCS is pivotal in quantum thermodynamics~\cite{Goold16,Vinjanampathy16},
adiabatic operations~\cite{Uchiyama14}, energy
backflow~\cite{Guarnieri16}, non-equilibrium fluctuation
relations~\cite{Riwar20} and studies of information~\cite{Utsumi17} and entropy
production~\cite{Nakajima17,Muller17}.

\emph{Perturbation expansions of $\G(t)$.
}
Our fixed-point relation~\eq{eq:fixed-point}
also allows challenges faced by perturbative calculations of
$\G$~\cite{Breuer01,Vacchini10,Timm11,Ferguson20} to be addressed.
Given some expansion of the memory kernel,
$\K=\K^{(1)}+\K^{(2)}+\ldots$, a corresponding series for $\G=\G^{(1)} +
\G^{(2)} + \ldots$ is obtained in terms of the \emph{memory
	kernel alone} without introducing any new formalism.
This is done by expanding~\Eq{eq:fixed-point} and organizing both sides
of the equation order by order. The first orders of $\G$ are then given by
\begin{subequations}
	\begin{align}
	\G^{(1)}(t,t_0) & = \int_{t_0}^t ds \, \K^{(1)}(t,s),
	\label{eq:expansion_G_1}
	\\
	\G^{(2)}(t,t_0) & = \int_{t_0}^t ds \, \K^{(2)}(t,s)
	\notag
	\\ &+
	i \int_{t_0}^t ds \int_{s}^t d\tau \, \K^{(1)}(t,s) \G^{(1)}(\tau,t_0).
	\label{eq:expansion_G_2}
	\end{align}\label{eq:expansion_G}\end{subequations}This reveals a recursive structure of perturbative $\G$ expansions.
Expanding in powers of the
system-environment coupling one recovers the approach of~\Ref{Breuer01}.
However, Eqs.~\eq{eq:expansion_G} is flexible and can also be applied to
expansions around a known \emph{dissipative} solution, which has
\emph{no} Hamiltonian formulation.
An interesting example is the $T=\infty$ solution of strongly
interacting, wide-band limit transport
models~\cite{Saptsov12,Saptsov14,Schulenborg16},
which leads to a powerful renormalized perturbation
theory~\cite{Saptsov14}.
This may be a step up for a renormalization
group~\cite{Schoeller09,Pletyukhov12a,Schoeller18,Lindner18,Lindner19}
treatment for time-local generators.

\emph{Nonperturbative semigroups and initial slippage.
}Our results can be used to gain detailed insight into approximation
strategies in the nonperturbative regime. In particular, the
superoperator $\S$ in \Eq{eq:inverse-laplace}-\eq{eq:S} is crucial for the \enquote{slippage of the
initial condition}, a well-known procedure aiming to improve Markovian
approximations~\cite{Geigenmuller83,Haake83,Haake85,Gaspard99,Yu00}.
In contrast to most previous works, we can derive definite statements
about the \emph{nonperturbative} quantity $\S$. This
has been explored in \Ref{Bruch20a} for the \emph{most favorable}
situation where a Markov approximation using the \emph{exact} Markovian
generator $\G(\infty)$ is improved upon using the \emph{exact} slippage
superoperator $\S$.
Even in this case subtle failures can arise, which are totally
unexpected within the \emph{time-local} formalism used to set up the slippage correction. Already for the example of the resonant level a dramatic breakdown occurs around apparently \enquote{innocent} isolated physical parameter points,
even though away
from these points it gives a considerably improved \emph{non-semigroup}
approximation.
However, this behavior can be clearly understood~\cite{Bruch20a} using the
connection~\eq{eq:sampling} to the time-nonlocal memory kernel
and applies generally to \emph{strongly interacting} fermionic
transport models \emph{far} from equilibrium.

\emph{Fixed-point analysis.
}Aside from these increasingly technical applications, perhaps the most
intriguing implication of our main result~\eq{eq:fixed-point} is the
possibility of analyzing the fixed-point iteration using ideas borrowed
from renormalization group transformations in statistical
physics~\cite{Wilson75} such as linearization around fixed points,
scaling, etc.
In the iteration
\eq{eq:fixed_point_decomposed} each new approximation $\G^{(n)}$ incorporates more of the memory integral over $\K$ in a step-wise fashion.
The apparent local stability of the fixed point of this discrete flow in
the functional superoperator space and its range of attraction must
somehow be related to physical retardation properties of the open system.
One could envisage comparing and perhaps even classifying open system
dynamics based on the nature of these discrete flows. 
}

\vspace{10pt}

\acknowledgments
We thank Y.-T. Lin, \refd{J. Schulenborg,} J. Splettstoesser, B. Vacchini, M. Pletyukhov, C. Timm, Y. Mokrousov and F. Lux for useful discussions.
K.N. and V.B. acknowledge support by the Deutsche Forschungsgemeinschaft (RTG 1995).

\appendix
\refb{\section{Iteration in the resonant level model\label{app:iteration_terminates}}

For the resonant level model both the stationary [\Sec{sec:stationary-iteration}] and transient  [\Sec{sec:time_dependent_iteration}] fixed-point iteration terminate at the first step when
starting from any zero-trace-preserving superoperator $X$, i.e., $\text{Tr} X \bullet=0$.
This can be seen by writing the time-nonlocal part of the kernel as~\cite{Saptsov14,Reimer19b}
\begin{align}
\K_N(t) = -i \Gamma \k(t) e^{-\Gamma t/2} G^+_+ G^+_-
\end{align}
in terms of fermionic \enquote{creation} \emph{super}operators~\cite{Saptsov12} satisfying $(G^+_\eta)^2=0$ and thus $G^+_{+} G^+_{-} G^+_{\eta} = 0$.
Using \enquote{second quantization} for superoperators~\cite{Saptsov14,Reimer19b} one expands any zero-trace-preserving superoperator $X$ in terms of products of superfields
and verifies that in each term $G^+_\eta$ stands on the far \emph{left}.
This implies
\begin{align}
\K_N(t-s) e^{i X (t-s)} = \K_N(t-s)
\end{align}
and by \Eq{eq:fixed_point_decomposed} the transient [\Eq{eq:G-simple}] and stationary [\Eq{eq:trivial}] iteration find the exact generator in a \emph{single} step.
Starting from an \emph{arbitrary} superoperator $X$,
the first iteration will \emph{make} it zero-trace by \Eq{eq:iteration-TP}
and by the above result the second iteration will be converged. }\section{Exact summation of the memory expansion of the time-nonlocal QME \label{app:memory}}

In Sec. \ref{sec:summing} of the text we mentioned that the memory expansions of \Refs{Contreras12,Karlewski14} are contained in our fixed-point relation~\eqref{eq:fixed-point}.
Here we give an explicit formula for \emph{all} terms. Moreover, we sum the series to a self-consistent form and recover our key results~\eq{eq:fixed-point} and \eq{eq:fixed-point-stationary}.

\emph{Memory-expansion.}
We essentially follow the approach of \Ref{Contreras12}
noting that we have verified that \Ref{Karlewski14} achieves exactly the same thing by manipulating partial integrations.
Both works start from the time-\emph{nonlocal} QME~\eqref{eq:qme-nonlocal-pi}
and construct the time-\emph{local} QME~\eq{eq:qme-local-pi}.
Importantly, \emph{no} weak coupling approximation is made in these works
but they do restrict attention to the stationary limit $t_0 \to -\infty$
by constructing the approximate time-local QME
$\tfrac{d}{dt}\Pi(t-t_0) \approx -i \G(\infty) \Pi(t-t_0)$.
This is the nonperturbative Markovian semigroup approximation discussed in \Sec{sec:approx}.
\Ref{Contreras12} considers only the leading memory-correction~\eq{eq:adiabatic}.
Here we make none of the mentioned assumptions
and specialize to the case of \Refs{Contreras12,Karlewski14} only at the end [\Eq{eq:actual}].

Thus, the summation of the memory expansion amounts to the construction of $\G(t,t_0)$ from $\K(t,s)$
such that we have
$\tfrac{d}{dt}\Pi(t,t_0) = -i \int_{t_0}^t ds \,  \K(t,s) \Pi(s,t_0)=-i \G(t,t_0) \Pi(t,t_0)$.
In the main text this was solved by exploiting the divisor,
$
\G(t,t_0) = \int_{t_0}^t ds \,  \K(t,s) \Pi(s,t|t_0)
$.
In our formulation, the approach taken in \Refs{Contreras12,Karlewski14} amounts to computing the divisor as
\begin{subequations}
	\begin{align}
	\Pi(s,t|t_0)
	&
	= \Pi(s,t_0) \Pi(t,t_0)^{-1}
	\\
	& =
	\sum_{k=0}^{\infty} \frac{1}{k!}(-1)^{k}(t-s)^k \F^k(t,t_0)
	\label{eq:memory-expansion}
\end{align}\end{subequations}by inserting the \emph{memory-expansion}
$\Pi(s,t_0) = \sum_{k}\tfrac{1}{k!}(-1)^{k}(t-s)^k \partial^{k}_t \Pi(t,t_0)$
of quantities in the past time $s$ around the present time $t>s$.
For example, \Eq{eq:adiabatic} discussed in \Ref{Contreras12} corresponds to the $k=0,1$ terms.
Here the superoperator-valued Taylor coefficients $\F^k(t,t_0)$
are the \emph{time-local generators of the $k$-th derivative} of the propagator:
\begin{align}
	\partial^{k}_t \Pi(t,t_0) = \F^k(t,t_0) \Pi(t,t_0)
	.
\end{align}
Written as 
$\F^k(t,t_0):= [\partial^{k}_t \Pi(t,t_0)]\Pi(t,t_0)^{-1}$
they are easily shown to obey the recursion relation
\begin{align}
	\F^k(t,t_0) = \partial_t \F^{k-1}(t,t_0) + \F^{k-1}(t,t_0) [-i \G(t,t_0)]
\end{align}
with starting condition $\F^0(t,t_0)=\ones$
giving, for instance,
\begin{align}
	\F^1(t,t_0)& =-i \G(t,t_0)
	\\
	\F^2(t,t_0)
	&
	=-i \partial_t \G(t,t_0)+[-i \G(t,t_0)]^2
	\notag\\
	\F^3(t,t_0)
	&
	=-i \partial_t^2 \G(t,t_0)
	+[-i \G(t,t_0)][-i \partial_t \G(t,t_0)]
	\notag
	\\
	&+2[-i \partial_t\G(t,t_0)] [-i \G(t,t_0)]
	+[-i \G(t,t_0)]^3
	\notag
\end{align}This suggests inserting the ansatz
\begin{align}
	\F^k(t,t_0)
	=
	& 
	\sum_{n=1}^{k}
	\sum_{p_1=0}^{k-n} \ldots \sum_{p_n=0}^{k-n} \delta_{k-n,p_1+ \ldots + p_n}	
	\notag \\
	& \times
	F^{n}_{p_1\ldots p_n} [-i \partial^{p_1}_t \G(t,t_0) ] \ldots [-i \partial^{p_n}_t \G(t,t_0) ]
	\label{eq:Fk}
\end{align}
and deriving the recursion relations for the coefficients
\begin{subequations}
	\begin{align}
	F^{n}_{p_1\ldots p_n}
	& =
	\sum_{j=1}^n 	F^{n}_{p_1\ldots (p_j-1) \ldots p_n}
	\quad \text{for } p_n \geq 1
	\label{eq:recursion1}
	,
\\
	F^{n}_{p_1\ldots p_{n-1}0}
	&
	=
	\sum_{j=1}^{n-1} 	F^{n}_{p_1\ldots (p_j-1) \ldots p_{n-1}0}
	+ F^{n-1}_{p_1 \ldots p_{n-1}}
	\label{eq:recursion2}
	.
	\end{align}\label{eq:recursion}\end{subequations}Together with the starting conditions $F^1_{0} = 1$
these define \emph{all} the coefficients of the memory expansion.
Construction of the general solution of the recursion equations~\eq{eq:recursion} is very cumbersome
and hides the elegant functional fixed-point relation.

\emph{Fixed-point equation.}
We now show that the result~\eq{eq:memory-expansion} with \eq{eq:Fk} 
equivalently follows  from our fixed-point relation~\eq{eq:fixed-point}
by inserting into \Eq{eq:solution-b} the memory expansion
$\G(s_i,t_0) = \sum_{p_i}\tfrac{1}{p_i!}(-t_i)^{p_i} \partial^{p_i}_t \G(t,t_0)$
and performing the nested integrations over variables $t_i = t-s_i$:
\begin{widetext}\begin{align}\Pi(s,t|t_0) &= \T_{\rightarrow} e^{ -\int^t_s d\tau [-i\G](\tau,t_0)}
=
\sum_{n=0}^\infty
(-1)^n
\underset{t > s_n > \ldots > s_1 > s}{\int d s_n \ldots d s_1}
[-i \G(s_1,t_0)] \ldots [-i \G(s_n,t_0)]
= \sum_{k=0}^\infty \frac{(-1)^k}{k!} (t-s)^k \F^k(t,t_0)
.
\label{eq:divisor1}
\end{align}\end{widetext}We obtain the \emph{explicit general} form of \emph{all} coefficients:
\begin{subequations}\begin{align}F^{n}_{p_1\ldots p_n}
	&
	=
	\frac{(n+\sum_i p_i)!}{\prod_i p_i! \prod_{i=0}^{n-1}[\sum_{j=0}^{i} p_{n-j}]}  
	\label{eq:F-coef-def}
	\\
	&
	=
	\prod_{i=1}^{n-1}
	\binom{p_{n-i} + \sum_{j=1}^{i-1} ( p_{n-j}+1) }
	{ p_{n-i} }
	\label{eq:F-coef-binom}
	\\
	&
	=
	\prod_{i=1}^{n-1}
	F^{2}_{p_{i-1}, p_i+\ldots+p_n + (n-i)}\refd{.}
	\label{eq:F-coef-F2}
\end{align}\label{eq:F-coef}\end{subequations}The factorization \eq{eq:F-coef-binom} into binomials shows that all coefficients are in fact integers. Using the form \eq{eq:F-coef-F2} one verifies~\footnote
	{One verifies \Eq{eq:recursion1}
	first for $n=2$, giving
	$F^2_{p_1p_2} = F^2_{(p_1-1)p_2} + F^2_{p_1(p_2-1)}$,
	and uses this relation for $n > 2$ to simplify the sum of the last two terms,
	$F^n_{p_1\ldots (p_{n-1}-1)p_n}+F^n_{p_1\ldots p_{n-1}(p_n-1)}$.
	Then one adds $F^n_{p_1\ldots (p_{n-2}-2)p_{n-1}p_n}$ and so forth.
	\Eq{eq:recursion2} follows as special case of \Eq{eq:recursion1} since
	$F^n_{p_1..p_{n-1}(-1)}=F^{n-1}_{p_1\ldots p_{n-1}}$.
	}
that the coefficients are indeed the solutions to the recursion relations~\eq{eq:recursion}.
With $\G(t,t_0) = \int_{t_0}^t ds \,  \K(t,s) \Pi(s,t|t_0)$ this establishes that the laborious determination of the coefficients and subsequent summation of the memory expansion~\eq{eq:memory-expansion} envisaged in \Refs{Contreras12,Karlewski14} ultimately leads to our general functional fixed-point equation~\eq{eq:fixed-point}.
Our derivation of this self-consistent equation in the main text circumvents all above complications by immediately identifying the divisor in \Eq{eq:generator-rel}. However, even if one is interested in generating memory expansions, our approach~\eq{eq:divisor1} via the divisor is far simpler.

Noting the special coefficient values $F^n_{0\ldots 0}=1$
we see that
$\F^k = [-i \G]^k + $~(terms involving at least one time-derivative of $\G$).
Thus in the stationary limit where $\lim_{t_0 \to -\infty}\partial^k_t \G(t,t_0)=0$
and $\lim_{t_0 \to -\infty} \G(t,t_0)=\G(\infty)$:
\begin{align}
\Pi(s-t|-\infty)
&
 = \sum_{k=0}^\infty \frac{(-1)^k}{k!} (t-s)^k [-i\G(\infty)]^k
\notag
\\
&
= e^{i\G(\infty) (t-s)}\refd{.}
\label{eq:actual}
\end{align}
Inserted into $\G(\infty) = \int_{0}^\infty ds \,  \K(t-s) \Pi(s-t|-\infty)$
we thus also directly recover our stationary fixed-point equation \eq{eq:fixed-point-stationary}
for time-translational systems, $\K(t,s)=\K(t-s)$ by explicit summation of the \emph{stationary} memory expansion. This is the specific expansion studied in \Refs{Contreras12,Karlewski14}.

\section{Relation of time-local generator and gradient/Moyal expansion of time-nonlocal QME\label{app:gradient}}

The memory expansion \eq{eq:memory-expansion} implies that the generator of the time-local QME
$\tfrac{d}{dt}\Pi(t,t_0) =-i \G(t,t_0) \Pi(t,t_0)$
may be written as a gradient expansion
\begin{align}
& \G(t,t_0)
=
\sum_{k=0}^\infty \frac{(-1)^k}{k!}
\Big[ \int_{t_0}^t ds \,  \K(t,s) (t-s)^k \Big] \F^k(t,t_0)
\notag
\\
&=
\sum_{k=0}^\infty \frac{1}{k!}
\Big[ \frac{\partial^k}{(-i \partial \E)^k} \hat{\K}(\E,t,t_0) \Big] \Big|_{\E=0}
\F^k(t,t_0)
\end{align}
with frequency derivatives of the Laplace-like integral transform 
$\hat{\K}(\E,t,t_0) := \int_{0}^{t-t_0} ds e^{i \E s} \K(t,t-s)$ of the memory kernel
(\enquote{finite-time Laplace transform}).
Since $\F^k(t,t_0):= [\partial^{k}_t \Pi(t,t_0)]\Pi(t,t_0)^{-1}=f(\G,\ldots,\partial_t^{k-1}\G)$
has no simple structure as function of $k$ [\Eq{eq:Fk}] it is not clear how the series can be summed, not even formally.
This reflects that it arises from the nontrivial anti-time-ordered exponential~\eq{eq:divisor1}.
If one instead considers its action on $\Pi(t,t_0)$,
\begin{align}
	&
	\tfrac{d}{dt}\Pi(t,t_0)
	\\
	&
	=
	-i
	\sum_{k=0}^\infty \frac{(-1)^k}{k!}
	\Big[ \int_{t_0}^t ds \,  \K(t,s) (t-s)^k \Big] \partial^{k}_t \Pi(t,t_0)
	\notag
	\\
	&
	=
	-i
	\sum_{k=0}^\infty \frac{1}{k!}
	\Big[ \frac{\partial^k}{(-i \partial \E)^k} \hat{\K}(\E,t,t_0) \Big] \Big|_{\E=0}
	\partial^{k}_t \Pi(t,t_0)
	\notag
	,
\end{align}
then the series can be \emph{formally} summed to give a nonlinear \emph{time-frequency-domain differential operator}. Its action on superoperator \emph{functions} of $t$ must coincide with the linear action of $\G(t,t_0)$ on the superoperator \emph{evaluated at}~$t$:
\begin{align}
	\hat{\K}(\E,t,t_0)
	e^{ i
		\frac{\overleftarrow{\partial}}{ \partial \E }
		\frac{\overrightarrow{\partial}}{ \partial t }
	}
	\Big|_{\E=0}
	\Pi(t,t_0)
	=
	\G(t,t_0) \Pi(t,t_0)\refd{.}
	\label{eq:moyal}	
\end{align}
Thus, $\G(t,t_0)$ here plays the role of a (superoperator-valued) eigenvalue of this \emph{time-domain} differential operator.
This differential operator is constructed as \emph{frequency-domain} differential operator acting to the \emph{left} on the memory kernel transform $\hat{\K}(\E,t,t_0)$.
The above follows the well-known Moyal approach~\cite{Moyal49,Groenewold46} to quantum physics of closed systems, where
one enforces locality at the price of introducing position- and momentum-space differential operators acting both to the right \emph{and to the left}.
Its extension to the \emph{time-nonlocal evolution} of open systems within the density-operator approach
is thus closely related to the time-convolutionless approach based on the time-local equation~\eq{eq:qme-local}.

Clearly, this \emph{formal} relation between the generator and the memory kernel is easily written down.
However, our functional fixed-point result~\eq{eq:fixed-point} goes beyond this by \emph{explicitly} expressing the action of the time-domain differential operator on the left hand side of~\Eq{eq:moyal}, evaluating $\partial^{k}_t \Pi(t,t_0) = \F^k(t,t_0) \Pi(t,t_0)$ [\Eq{eq:Fk}],
and summing the series to an anti-time-ordered exponential in terms of $\G(t,t_0)$.
This is demonstrated by \Eq{eq:divisor1} read in reverse order.
As the main text shows, this makes the fixed-point relation a powerful analytical and numerical tool.

For time-translational systems, $\K(t,s)=\K(t-s)$,
taking the stationary limit leads to the simplification $\lim_{t_0 \to -\infty} \F^k(t,t_0) = (-i \G(\infty))^k$, giving
\begin{align}
\G(\infty)
& =
\sum_{k=0}^\infty \frac{(-1)^k}{k!}
\Big[ \int_{-\infty}^t ds \,  \K(t-s) (t-s)^k \Big] (-i\G(\infty))^k
\notag
\\
& =
\sum_{k=0}^\infty \frac{1}{k!}
\Big[ \frac{\partial^k}{\partial \E^k} \hat{\K}(\E) \Big] \Big|_{\E=0}
\G(\infty)^k
\end{align}with frequency-derivatives \refall{of the}
Laplace-transformed memory kernel $\hat{\K}(\E) = \int_{0}^\infty ds \,  \K(s) e^{is\E}$.
In this case, the gradient expansion can be summed to give an alternative expression for our stationary fixed point relation~\eq{eq:fixed-point-stationary}:
\begin{align}
	\hat{\K}(\E)
	e^{
		\frac{\overleftarrow{\partial}}{ \partial \E } \G(\infty)
	}
	\big|_{\E=0}
	=
	\G(\infty)\refd{.}
	\label{eq:shift}
\end{align}This gives a nonlinear differential operator acting to the left on superoperator functions of $\E$
and is a mere \emph{formal} expression of our stationary fixed-point equation~\eq{eq:fixed-point-stationary},
$\hat{\K}(\G(\infty))
:=
\int_{-\infty}^t ds \K(t-s)
e^{i\G(\infty) (t-s)}$.
Equation~\eq{eq:shift} extends the shift property for ordinary Laplace transforms,
$e^{\frac{\partial}{\partial \E} \Delta} \hat{f}(\E)=\hat{f}(\E + \Delta)$,
to our result~\eq{eq:fixed-point-stationary-b} with superoperator-valued frequency argument $\Delta=\G(\infty)$.
To ensure that the memory kernel generates a trace-preserving evolution [\Eq{eq:iteration-TP}]
the frequency derivatives must \emph{stand on the right} and therefore needs to \emph{act to the left}
to accomplish the shift.

 {}

\end{document}